\documentclass[fleqn,usenatbib]{mnras}

\usepackage[T1]{fontenc}

\DeclareRobustCommand{\VAN}[3]{#2}
\let\VANthebibliography\thebibliography
\def\thebibliography{\DeclareRobustCommand{\VAN}[3]{##3}\VANthebibliography}

\usepackage{graphicx}	% Including figure files
\usepackage{amsmath}	% Advanced maths commands

\usepackage{soul}       % strike out text with \st{.......}
\title[Type Ia Supernovae SN 2013bz, PSN J0910+5003 and ASASSN-16ex]{Type Ia supernovae
SN 2013bz, PSN J0910+5003 and ASASSN-16ex: similar to 09dc--like? }

\author[S. Tiwari et al.]{S. Tiwari,$^{1}$
N. K. Chakradhari,$^{1,2}$
D. K. Sahu,$^3$\thanks{E-mail: dks@iiap.res.in} 
G. C. Anupama,$^3$
B. Kumar$^4$
and K. R. Sahu$^{1}$
\\
$^1$School of Studies in Physics \& Astrophysics, Pt. Ravishankar Shukla University, Raipur 492010, India\\
$^2$Centre for Mega Projects in Multiwavelength Astronomy, Pt. Ravishankar Shukla University, Raipur 492010, India\\
$^3$Indian Institute of Astrophysics, Koramangala, Bangalore 560034, India\\
$^4$Aryabhatta Research Institute of observational sciencES (ARIES), Manora Peak, Nainital 263001, India
}

\date{Accepted XXX. Received YYY; in original form ZZZ}

\pubyear{2023}

\begin{document}
\label{firstpage}
\pagerange{\pageref{firstpage}--\pageref{lastpage}}
\maketitle

% Abstract of the paper
\begin{abstract}
We present optical photometric and spectroscopic studies of three supernovae (SNe) SN 2013bz, PSN J0910+5003 and ASASSN-16ex.   UV-optical photometric data of ASASSN-16ex obtained with  Swift-UVOT  are also analyzed. These objects  were initially classified as 09dc--like type Ia SNe. The decline rate parameters ($\Delta m_{15}(B)_{true}$) are derived as 0.92 $\pm$ 0.04 (SN 2013bz), 0.70 $\pm$ 0.05 (PSN J0910+5003) and 0.73 $\pm$ 0.03 (ASASSN-16ex). The estimated $B$ band absolute magnitudes at maximum:  $-$19.61 $\pm$ 0.20\,mag for SN 2013bz, $-$19.44 $\pm$ 0.20\,mag for PSN J0910+5003 and $-$19.78 $\pm$ 0.20\,mag for ASASSN-16ex indicate that all the three objects are relatively bright. The peak bolometric luminosities for these objects are derived as $\log L_\text{bol}^\text{max} =$ 43.38 $\pm$ 0.07\,erg\,s$^{-1}$, 43.26 $\pm$ 0.07\,erg\,s$^{-1}$ and 43.40 $\pm$ 0.06\,erg\,s$^{-1}$, respectively. 
The spectral and  velocity evolution of SN 2013bz is similar to a normal SN Ia, hence  it appears to be a luminous, normal type Ia supernova. On the other hand, the light curves of  PSN J0910+5003 and ASASSN-16ex are broad and  exhibit properties similar to 09dc--like SNe Ia. Their  spectroscopic evolution  shows similarity with 09dc--like SNe, strong C\,{\sc ii} lines  are  seen in the pre-maximum spectra of these two events. Their photospheric velocity evolution is similar to SN 2006gz. Further, in the UV bands, ASASSN-16ex is very blue like other 09dc--like SNe Ia. 
\end{abstract}

% Select between one and six entries from the list of approved keywords.
% Don't make up new ones.

%%%%%%%%%%%%%%%%%%%%%%%%%%%%%%%%%%%%%%%%%%%%%%%%%%

%%%%%%%%%%%%%%%%% BODY OF PAPER %%%%%%%%%%%%%%%%%%

\section{Introduction}
\label{sec_intro}
Thermonuclear Supernovae are an important class of supernovae (SNe), whose progenitors are low-mass stars found in elliptical as well as spiral galaxies. They are commonly known as Type-Ia SNe (SNe Ia) and populate the brighter side of the luminosity distribution of SNe. Most SNe Ia, referred to as `normal SNe Ia', display uniform spectral and light-curve properties. 
Their luminosity is correlated with the width of their light curve \citep{phil93,phil99} and hence are considered standardizable candles. This uniformity and high luminosity, make them a vital  probe for studying cosmic evolution \citep{ries98,perl99}. SNe Ia are the primary source of Iron Group Elements (IGEs), hence play an important role in enriching the Inter-Stellar Medium (ISM) with IGEs  (\citealt{matt86,matt09}; \citealt*{nomo13}). 

Our understanding of the progenitor and explosion mechanism giving rise to these events still needs to be completed. From the theoretical and observational work, it is inferred that thermonuclear disruption of a carbon-oxygen (C/O) White Dwarf (WD) in a binary system results in a type Ia explosion (\citealt{hoyl60}, see \citealt*{maoz14,jhas19} for reviews). There are two possible progenitor models suggested for a WD to explode. In the first one, a WD accretes matter from a non-degenerate star, known as the single degenerate (SD) model \citep{whel73}. In the double degenerate (DD) model, the explosion results from the merger of two WDs \citep{iben84,webb84}. Most SNe Ia are considered to be an explosion of Chandrasekhar mass WD \citep{mazz07} via delayed detonation \citep{khok91}. However, if the accumulated material is He-rich, the explosion can occur at a sub-Chandrasekhar mass through double-detonation. With sufficiently rapid He accretion on the surface of a C/O WD, a detonation is first initiated within the helium layer. The emanating shock wave  propagates through the WD and triggers carbon detonation at the center of the WD \citep{woos94,woos11,ruit14,tani18}. The donor star could be either a non-degenerate He star (SD channel), another C/O WD with He in the outer layer or a He WD (DD channel). This mechanism can explain normal and fast declining SNe Ia of different brightness distributions \citep{pakm13}.  Currently, it is difficult  to identify which SN results from which channel \citep[see][for reviews]{wang18,livi18,soke19,ruit20}.

With the increasing number of well-studied SNe Ia, it became clear that there is a considerable spread in the luminosity of SNe Ia. There are objects populating  both the higher and lower luminosity end of normal objects \citep{liwe11_rate}. Some have extreme properties that are significantly different from the normal class \citep{taub17}. 

A small group of extreme/peculiar objects (e.g., SN 2003fg: \citealt{howe06}, SN 2006gz: \citealt{hick07,maed09}, SN 2007if: \citealt{scal10,yuan10}, SN 2009dc: \citealt{yama09,silv11,taub11})
are known to have broad light curves and generally have a brighter peak luminosity than predicted by their decline rate \citep{phil99}. In most of these, the ejecta mass inferred through simple analytic modelling, and hence the progenitor mass is found to be higher than the Chandrasekhar mass,  
making them to be termed as super-Chandrasekhar SNe Ia or 03fg/09dc--like SNe Ia. They are UV blue and bright in the early phase. 
Near maximum spectra of 09dc--like SNe are similar to the normal events, characterized by intermediate-mass elements (IMEs) but with relatively narrow/weak spectral features. 
The photospheric velocity is found to be low, with a slow evolution.
The presence of unburned carbon in the ejecta is often seen in the pre-maximum spectral sequence.  

The light-curve and spectral evolution of SN 2012dn was similar to 09dc--like objects. However, it had weak carbon, relatively strong features from IMEs, and was fainter compared to other  09dc--like SNe Ia 
\citep{brow14,chak14,parr16,yama16,taub19}. Now it appears that even this group (with some new members, e.g., ASASSN-15pz, LSQ14fmg, ASASSN-15hy and SN 2020esm) exhibits diverse observational properties \citep{chen19,hsia20,luj21,asha21,dimi22}. 
The luminous type Ia SN 2020hvf \citep{jian21}, SN 2021zny \citep{dimi23} and SN 2022ilv \citep{sriv23} were also found to be similar to 09dc--like SNe having broad light curves and  strong carbon features in the early spectra. Early excess emission was seen in these SNe within a few hours of the explosion.  

In this work, we present a detailed photometric and spectroscopic study of three SNe Ia: SN 2013bz, PSN J0910+5003 and ASASSN-16ex. They were initially classified as 09dc--like SNe Ia. 

\begin{figure}
\centering
\includegraphics[width=0.846\columnwidth]{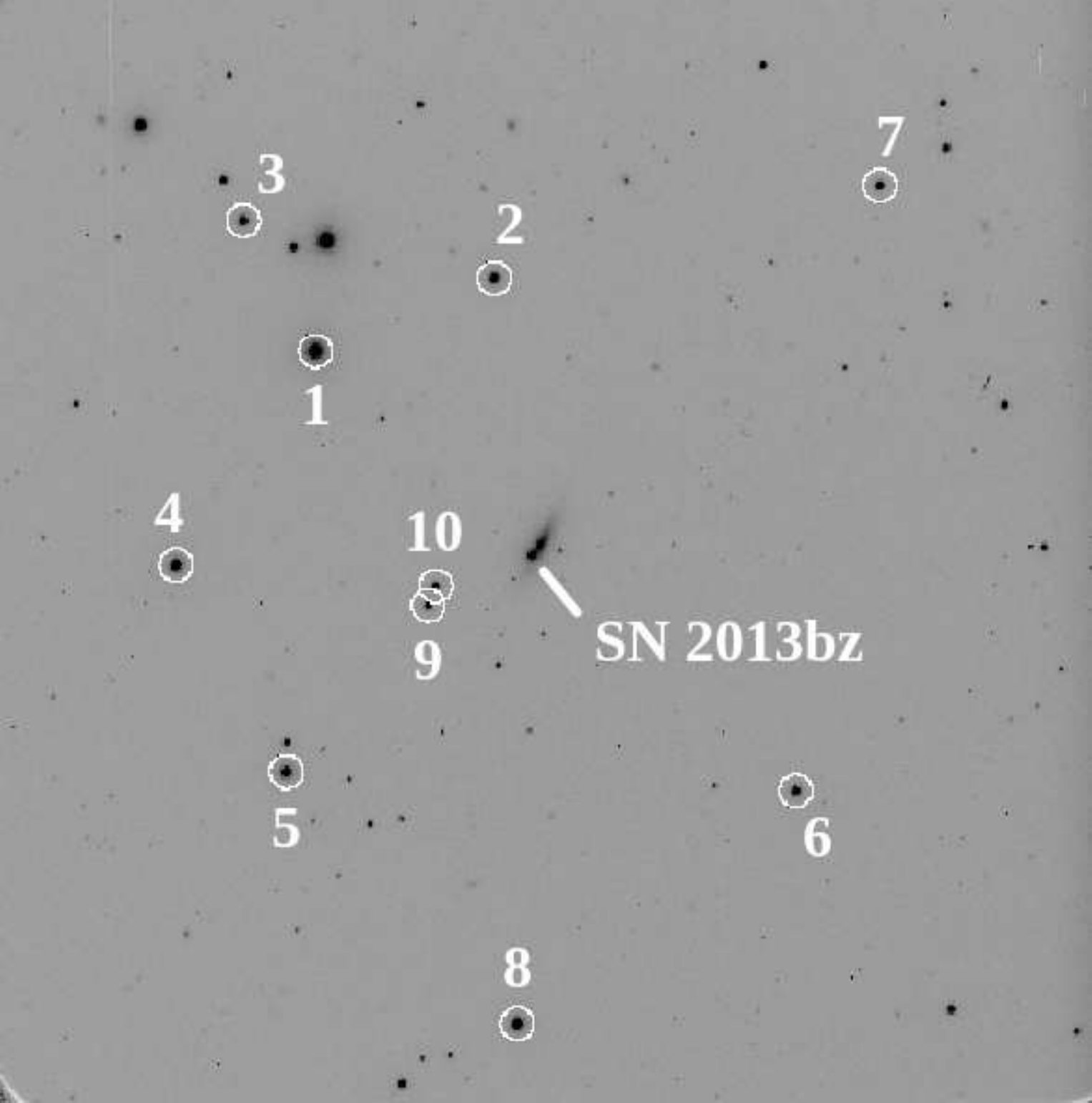}
\includegraphics[width=0.846\columnwidth]{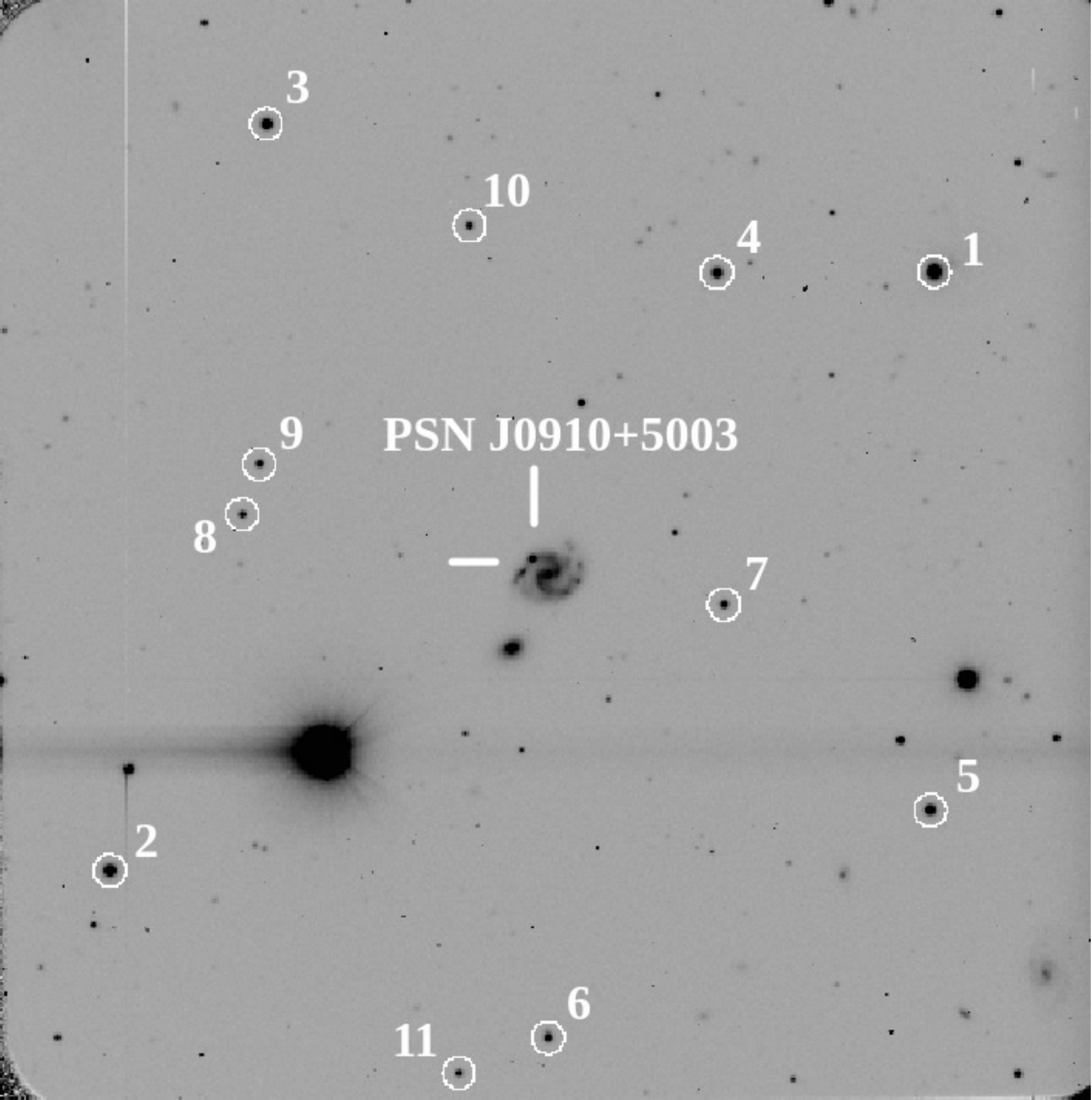}
\includegraphics[width=0.846\columnwidth]{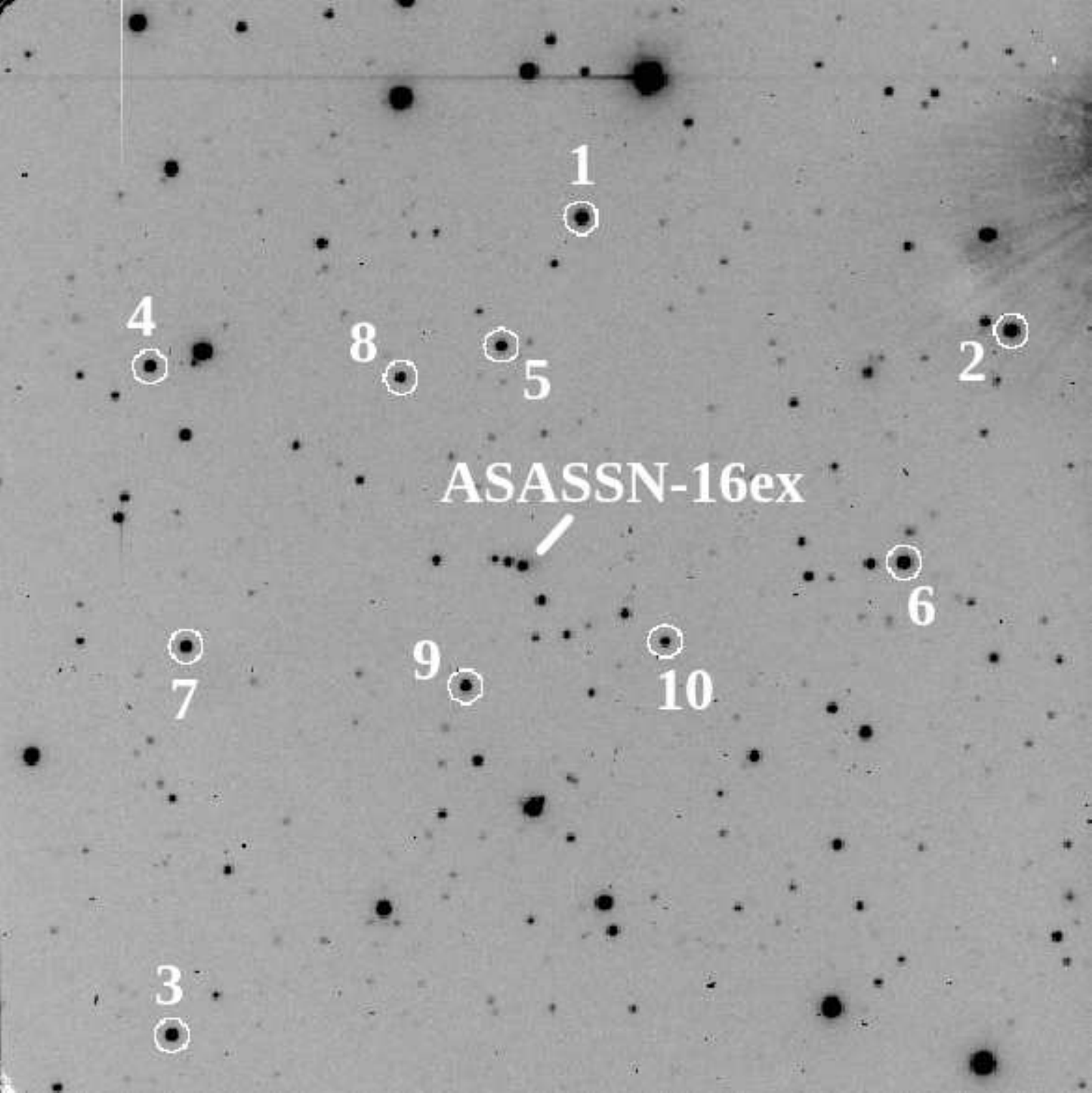}
\caption{Identification chart for SNe 2013bz ({\it top}), PSN J0910+5003 ({\it middle}) and ASASSN-16ex ({\it bottom}). The stars used as secondary standards are marked. North is up and east to the left. The field of view is 10 $\times$ 10 arcmin$^2$ each. Images were obtained using 2-m HCT, IAO, Hanle, India.}
\label{figsn_field}
\end{figure}

The Catalina Real-time Transient Survey discovered SN 2013bz on April 21.28 (time is always in {\sc ut}), 2013 and classified it as a normal type Ia SN around the maximum. However, noted that it could  also be an overluminous event  \citep{howe13}.  \citet{ochn13} reported that the spectrum obtained on May 4.96, 2013, matched well with SNe 2006gz/09dc. SN was located at RA = 13:26:51.32, DEC = -10:01:32.2 (J2000), 7.5 arcsec East and  5.8 arcsec South to the nucleus of host galaxy PGC 170248 (Fig.~\ref{figsn_field}, {\it top}).

The Italian SNe Search Project discovered PSN J0910+5003 on November 8.17, 2015 \citep*{ciab15}. The object was located at RA = 09:10:08.85,  DEC = 50:03:39.6 (J2000), 9 arcsec East and 8 arcsec North of the centre of host galaxy UGC 4812 (Fig.~\ref{figsn_field}, {\it middle}). A spectrum obtained on November 09.13, 2015, matched well with  SN 2009dc about ten days before maximum light \citep{toma15}. 

All Sky Automated Survey for SNe discovered ASASSN-16ex (SN 2016ccj) on May 03.43, 2016 \citep{kiyo16}. The SN was located at RA = 17:10:23.92, DEC= 26:23:47.89 (J2000), 1.9 arcsec South and 3.8 arcsec East from the centre of the host galaxy SDSS J171023.63+262350.3 (Fig.~\ref{figsn_field}, {\it bottom}). 
Spectra obtained by \citet{toma16,pias16} matched well with SNe 2009dc/06gz, about 10--12 days before maximum. Prominent C\,{\sc ii} lines (6580\,\AA\ and 7234\,\AA)  were also detected in the spectra.  The UV colours and absolute magnitudes of ASASSN-16ex were found to be blue and bright similar to 09dc--like events \citep{brow16}.

The layout of this paper is as follows. 
Observations and data reduction techniques are presented in Section \ref{sec_observation}. 
The light curves and colour curves are discussed in Section \ref{sec_light_curve}. 
Physical parameters of the SNe, e.g., peak absolute magnitude, bolometric luminosity and mass of $^{56}$Ni, etc., are estimated in Section \ref{sec_abs_bol_luminosity}. 
The spectral evolution and comparison with other well-studied SNe Ia are presented in Section \ref{spectral_evolution}. 
The results are discussed and summarized in Section \ref{sec_summary}. Throughout this paper, we adopt the Vega magnitude system.

\section{OBSERVATIONS AND DATA REDUCTION}
\label{sec_observation}
\subsection{Imaging}

SN 2013bz, PSN J0910+5003 and ASASSN-16ex  were monitored using the 2-m Himalayan Chandra Telescope (HCT) of the Indian Astronomical Observatory (IAO\footnote{\url{https://www.iiap.res.in/?q=iao.htm}}), Hanle, India. 
Observations were carried out in Bessell’s $UBVRI$ bands using the Himalaya Faint Object Spectrograph Camera (HFOSC).  

Photometric monitoring of SN 2013bz started on May 02, 2013 and continued until July 04, 2013.  Standard star fields PG 1633+099, PG 2213-006, PG 1525-71 and PG 1323-086 from \citet{land92} were observed
on six photometric nights
(May 02, May 14, May 21, May 30, June 04 and July 04, 2013) to calibrate a sequence of secondary
standards in the SN field.

Photometric observations of PSN J0910+5003 were carried out from November 10, 2015, to March 13, 2016. Standard star fields PG 0231+051, PG 0918+029, PG 2213-006, PG 1047+003 and PG 0942-029 were observed on five photometric nights (November 17, November 20, November 28, 2015; December 01, 2020; and March 01, 2021).  

Photometric monitoring of ASASSN-16ex began on May 07, 2016, and continued till September 17, 2016. Standard star fields PG 1633+099, SA 110, PG 2213-006, PG 0231+051 and PG 2331+055 were observed
on seven photometric nights (May 14, June 03, June 18, June 26, July 06, July 31 and October 04, 2016). Several bias frames  and twilight flats necessary for pre-processing were also obtained during each observation. 

The observed data  were processed in a standard manner using the Image Reduction and Analysis Facility ({\sc iraf}\footnote{\url{https://iraf-community.github.io/}}) package. The observed Landolt standards were used to calibrate a sequence of secondary standards in the SNe field, following the procedures of \citet{chak14}; \citet*{chak19}. 
The mean $UBVRI$ magnitudes of secondary standards  in the field of SN 2013bz, PSN J0910+5003 and ASASSN-16ex (marked in Fig.~\ref{figsn_field}) are listed in Table~\ref{tab_sec_std}. 

\begin{table*}
\caption{Magnitudes of secondary standards (marked in Fig.~\ref{figsn_field}) in the field of SN 2013bz, PSN J0910+5003 and ASASSN-16ex.}
\small
\centering
\begin{tabular}{@{}lccccc@{}}
\hline
ID & U  & B & V &  R & I\\
\hline
\multicolumn{6}{@{}c}{\bf SN 2013bz \vspace{1ex}}\\
1 &14.34 $\pm$ 0.02 & 14.22 $\pm$ 0.03 & 13.51 $\pm$ 0.02 & 13.20 $\pm$ 0.03 & 12.81 $\pm$ 0.02\\	
2 &15.64 $\pm$ 0.02 & 15.61 $\pm$ 0.03 & 14.92 $\pm$ 0.02 & 14.60 $\pm$ 0.02 & 14.18 $\pm$ 0.02\\  
3 &15.60 $\pm$ 0.02 & 15.78 $\pm$ 0.03 & 15.16 $\pm$ 0.02 & 14.85 $\pm$ 0.02 & 14.43 $\pm$ 0.01\\  
4 &14.92 $\pm$ 0.01 & 15.08 $\pm$ 0.03 & 14.47 $\pm$ 0.02 & 14.18 $\pm$ 0.02 & 13.78 $\pm$ 0.02\\  
5 &15.20 $\pm$ 0.01 & 15.36 $\pm$ 0.02 & 14.75 $\pm$ 0.02 & 14.46 $\pm$ 0.02 & 14.06 $\pm$ 0.01\\  
6 &16.51 $\pm$ 0.03 & 16.28 $\pm$ 0.02 & 15.52 $\pm$ 0.02 & 15.17 $\pm$ 0.02 & 14.73 $\pm$ 0.02\\  
7 &17.14 $\pm$ 0.05 & 16.63 $\pm$ 0.02 & 15.73 $\pm$ 0.02 & 15.30 $\pm$ 0.02 & 14.77 $\pm$ 0.01\\  
8 &15.66 $\pm$ 0.03 & 15.36 $\pm$ 0.02 & 14.54 $\pm$ 0.02 & 14.19 $\pm$ 0.03 & 13.71 $\pm$ 0.02\\  
9 &17.65 $\pm$ 0.05 & 16.92 $\pm$ 0.03 & 15.96 $\pm$ 0.02 & 15.50 $\pm$ 0.02 & 14.98 $\pm$ 0.01\\  
10&16.76 $\pm$ 0.03 & 16.74 $\pm$ 0.02 & 16.05 $\pm$ 0.02 & 15.73 $\pm$ 0.02 & 15.31 $\pm$ 0.02\\  
\multicolumn{6}{@{}c}{ \vspace{-2ex}}\\
\multicolumn{6}{@{}c}{\bf PSN J0910+5003 \vspace{1ex}}\\
1 &14.43 $\pm$ 0.04 & 14.51 $\pm$ 0.04 & 13.94 $\pm$ 0.02 & 13.59 $\pm$ 0.02 & 13.35 $\pm$ 0.02\\	
2 &16.11 $\pm$ 0.04 & 15.80 $\pm$ 0.02 & 15.02 $\pm$ 0.01 & 14.56 $\pm$ 0.02 & 14.17 $\pm$ 0.04\\	
3 &17.58 $\pm$ 0.01 & 16.50 $\pm$ 0.04 & 15.39 $\pm$ 0.02 & 14.73 $\pm$ 0.02 & 14.16 $\pm$ 0.01\\	
4 &17.44 $\pm$ 0.01 & 16.61 $\pm$ 0.04 & 15.62 $\pm$ 0.03 & 15.05 $\pm$ 0.02 & 14.57 $\pm$ 0.02\\	
5 &17.02 $\pm$ 0.04 & 16.49 $\pm$ 0.04 & 15.62 $\pm$ 0.02 & 15.12 $\pm$ 0.02 & 14.66 $\pm$ 0.03\\	
6 &19.10 $\pm$ 0.05 & 18.06 $\pm$ 0.04 & 16.92 $\pm$ 0.03 & 16.29 $\pm$ 0.03 & 15.73 $\pm$ 0.03\\	
7 &19.57 $\pm$ 0.08 & 18.54 $\pm$ 0.03 & 17.04 $\pm$ 0.03 & 16.12 $\pm$ 0.01 & 15.25 $\pm$ 0.03\\	
8 &18.65 $\pm$ 0.01 & 18.19 $\pm$ 0.04 & 17.32 $\pm$ 0.02 & 16.86 $\pm$ 0.04 & 16.44 $\pm$ 0.04\\	
9 &19.90 $\pm$ 0.02 & 18.92 $\pm$ 0.03 & 17.45 $\pm$ 0.02 & 16.36 $\pm$ 0.03 & 15.01 $\pm$ 0.03\\	
10&19.21 $\pm$ 0.02 & 18.52 $\pm$ 0.03 & 17.51 $\pm$ 0.03 & 16.93 $\pm$ 0.03 & 16.38 $\pm$ 0.03\\	
11&19.42 $\pm$ 0.01 & 18.54 $\pm$ 0.08 & 17.51 $\pm$ 0.04 & 16.87 $\pm$ 0.04 & 16.27 $\pm$ 0.04\\	
\multicolumn{6}{@{}c}{ \vspace{-2ex}}\\
\multicolumn{6}{@{}c}{\bf ASASSN-16ex \vspace{1ex}}\\
1 &15.91 $\pm$ 0.02 & 15.89 $\pm$ 0.03 & 15.21 $\pm$ 0.02 & 14.84 $\pm$ 0.02 & 14.50 $\pm$ 0.02\\
2 &16.31 $\pm$ 0.03 & 16.04 $\pm$ 0.04 & 15.22 $\pm$ 0.02 & 14.74 $\pm$ 0.02 & 14.29 $\pm$ 0.03\\
3 &16.36 $\pm$ 0.02 & 16.37 $\pm$ 0.04 & 15.70 $\pm$ 0.02 & 15.33 $\pm$ 0.03 & 14.98 $\pm$ 0.03\\
4 &16.73 $\pm$ 0.04 & 16.70 $\pm$ 0.03 & 16.01 $\pm$ 0.02 & 15.64 $\pm$ 0.02 & 15.28 $\pm$ 0.03\\
5 &16.88 $\pm$ 0.04 & 16.88 $\pm$ 0.03 & 16.19 $\pm$ 0.02 & 15.81 $\pm$ 0.01 & 15.44 $\pm$ 0.02\\
6 &16.89 $\pm$ 0.02 & 16.30 $\pm$ 0.04 & 15.38 $\pm$ 0.02 & 14.85 $\pm$ 0.01 & 14.36 $\pm$ 0.02\\
7 &17.02 $\pm$ 0.05 & 16.89 $\pm$ 0.04 & 16.13 $\pm$ 0.02 & 15.72 $\pm$ 0.02 & 15.33 $\pm$ 0.02\\
8 &17.00 $\pm$ 0.05 & 17.35 $\pm$ 0.02 & 16.79 $\pm$ 0.02 & 16.44 $\pm$ 0.02 & 16.09 $\pm$ 0.02\\
9 &17.15 $\pm$ 0.05 & 17.09 $\pm$ 0.04 & 16.41 $\pm$ 0.02 & 16.03 $\pm$ 0.02 & 15.65 $\pm$ 0.02\\
10&17.36 $\pm$ 0.03 & 17.70 $\pm$ 0.03 & 17.16 $\pm$ 0.02 & 16.83 $\pm$ 0.01 & 16.49 $\pm$ 0.03\\
\hline
\multicolumn{6}{@{}l}{All magnitudes are in the Vega system.}
\end{tabular}
\label{tab_sec_std}
\end{table*}

The host galaxy background significantly contaminated SN 2013bz and PSN J0910+5003. 
Template subtraction photometry was performed for these SNe to remove contribution from the host on the SN flux.
Deep  $UBVRI$ template frames of PGC 170248 and UGC 4812 fields were obtained in good seeing condition with the same instrumental setup on March 15, 2016, and March 01, 2021, respectively. Template image subtraction was performed in a standard manner using {\sc iraf} {\it cl}-script. 
In the template subtracted frame, aperture photometry was performed on the SN and calibrated differentially with respect to the secondary standards. The estimated magnitudes of SN 2013bz and PSN J0910+5003 are listed in Table~\ref{tab_sn_mag}.

For ASASSN-16ex, profile-fitting photometry was used. The fitting radius was chosen close to the FWHM of the stellar profile. A correction factor was obtained from the difference between the aperture and profile fitting magnitudes using bright field stars. This correction factor was applied to the SN magnitude derived using profile fitting, which was then calibrated differentially with respect to the secondary standards in the field. The derived magnitudes of ASASSN-16ex are listed in Table~\ref{tab_sn_mag}. The magnitudes presented in this paper are not K-corrected.

\begin{table*}
\caption{Optical $UBVRI$ photometry of SN 2013bz, PSN J0910+5003and ASASSN-16ex with HCT.}
\small
\centering
\begin{tabular}{@{}lcrccccc@{}}
\hline
 Date & JD & Phase$^a$ & U & B & V & R & I\\
\hline
\multicolumn{8}{@{}c}{\bf SN 2013bz \vspace{1ex}}\\
02/05/2013& 245\,6415.22 &5.72  & 15.49 $\pm$ 0.02 & 15.90 $\pm$ 0.01 & 15.55 $\pm$ 0.01& 15.52 $\pm$ 0.01& 15.72 $\pm$ 0.01\\  	 
03/05/2013& 245\,6416.16 &6.66  &                  & 15.93 $\pm$ 0.01 & 15.54 $\pm$ 0.01& 15.54 $\pm$ 0.01& 15.79 $\pm$ 0.03\\  	
05/05/2013& 245\,6418.18 &8.68  &                  & 16.09 $\pm$ 0.01 & 15.65 $\pm$ 0.01& 15.66 $\pm$ 0.01& 15.94 $\pm$ 0.02\\    
06/05/2013& 245\,6419.22 &9.72  & 15.77 $\pm$ 0.01 & 16.11 $\pm$ 0.01 & 15.71 $\pm$ 0.01& 15.74 $\pm$ 0.02& 16.01 $\pm$ 0.02\\    
08/05/2013& 245\,6421.24 &11.74 & 15.94 $\pm$ 0.02 & 16.30 $\pm$ 0.02 & 15.83 $\pm$ 0.01& 15.96 $\pm$ 0.01& 16.19 $\pm$ 0.01\\  	
14/05/2013& 245\,6427.13 &17.63 &                  & 16.91 $\pm$ 0.01 & 16.17 $\pm$ 0.01& 16.19 $\pm$ 0.01& 16.19 $\pm$ 0.02\\  	
18/05/2013& 245\,6431.21 &21.71 &                  & 17.36 $\pm$ 0.01 & 16.32 $\pm$ 0.01& 16.17 $\pm$ 0.01& 16.03 $\pm$ 0.01\\  	
20/05/2013& 245\,6433.21 &23.71 & 17.77 $\pm$ 0.17 & 17.56 $\pm$ 0.02 & 16.40 $\pm$ 0.01& 16.10 $\pm$ 0.02& 16.01 $\pm$ 0.01\\  	
21/05/2013& 245\,6434.26 &24.76 &                  & 17.85 $\pm$ 0.02 & 16.55 $\pm$ 0.02& 16.34 $\pm$ 0.05& 16.16 $\pm$ 0.08\\    
27/05/2013& 245\,6440.18 &30.68 &                  & 18.25 $\pm$ 0.02 & 16.80 $\pm$ 0.02& 16.44 $\pm$ 0.01& 16.08 $\pm$ 0.02\\    
28/05/2013& 245\,6441.23 &31.73 &                  &                  & 16.87 $\pm$ 0.01& 16.53 $\pm$ 0.02& 16.09 $\pm$ 0.01\\  	
29/05/2013& 245\,6442.22 &32.72 & 18.37 $\pm$ 0.05 & 18.34 $\pm$ 0.01 & 16.89 $\pm$ 0.01& 16.52 $\pm$ 0.01& 16.06 $\pm$ 0.02\\  	
30/05/2013& 245\,6443.23 &33.73 & 18.42 $\pm$ 0.03 & 18.40 $\pm$ 0.01 & 16.96 $\pm$ 0.01& 16.57 $\pm$ 0.01& 16.14 $\pm$ 0.02\\    
31/05/2013& 245\,6444.19 &34.69 & 18.40 $\pm$ 0.04 & 18.43 $\pm$ 0.01 & 16.97 $\pm$ 0.01& 16.66 $\pm$ 0.01& 16.28 $\pm$ 0.02\\    
04/06/2013& 245\,6448.21 &38.71 & 18.69 $\pm$ 0.04 & 18.66 $\pm$ 0.01 & 17.30 $\pm$ 0.01& 16.89 $\pm$ 0.02& 16.47 $\pm$ 0.02\\     
13/06/2013& 245\,6457.23 &47.73 &                  & 18.72 $\pm$ 0.03 & 17.57 $\pm$ 0.05& 17.35 $\pm$ 0.02& 16.94 $\pm$ 0.03\\    
04/07/2013& 245\,6478.17 &68.67 & 18.98 $\pm$ 0.05 & 19.17 $\pm$ 0.02 & 18.19 $\pm$ 0.01& 18.13 $\pm$ 0.02& 18.15 $\pm$ 0.03\\  	 
\multicolumn{8}{@{}c}{ \vspace{-2ex}}\\
\multicolumn{8}{@{}c}{\bf PSN J0910+5003 \vspace{1ex}}\\
10/11/2015& 245\,7337.46 & $-$9.54&                  & 17.71 $\pm$ 0.04& 17.65 $\pm$ 0.03& 17.54 $\pm$ 0.02& 17.55 $\pm$ 0.02\\	
17/11/2015& 245\,7344.38 & $-$2.62& 16.70 $\pm$ 0.11 & 17.16 $\pm$ 0.03& 17.02 $\pm$ 0.02& 16.89 $\pm$ 0.02& 16.93 $\pm$ 0.02\\        
19/11/2015& 245\,7346.49 & $-$0.51&                  & 17.15 $\pm$ 0.05& 16.99 $\pm$ 0.04& 16.82 $\pm$ 0.06& 16.94 $\pm$ 0.03\\        
20/11/2015& 245\,7347.42 &    0.42&                  & 17.12 $\pm$ 0.03& 16.95 $\pm$ 0.02& 16.79 $\pm$ 0.02& 16.97 $\pm$ 0.02\\        
22/11/2015& 245\,7349.45 &    2.45& 16.69 $\pm$ 0.04 & 17.18 $\pm$ 0.07& 16.92 $\pm$ 0.03& 16.71 $\pm$ 0.02& 16.80 $\pm$ 0.05\\        
27/11/2015& 245\,7354.44 &    7.44& 17.07 $\pm$ 0.17 & 17.32 $\pm$ 0.04& 16.91 $\pm$ 0.04& 16.73 $\pm$ 0.04& 16.89 $\pm$ 0.04\\	
28/11/2015& 245\,7355.42 &    8.42&                  & 17.49 $\pm$ 0.08& 16.95 $\pm$ 0.03& 16.79 $\pm$ 0.10& 16.80 $\pm$ 0.04\\    	
01/12/2015& 245\,7358.46 &   11.46&                  & 17.57 $\pm$ 0.04& 17.03 $\pm$ 0.03& 16.80 $\pm$ 0.02& 16.91 $\pm$ 0.04\\	
09/12/2015& 245\,7366.39 &   19.39&                  & 18.13 $\pm$ 0.03& 17.31 $\pm$ 0.02& 17.06 $\pm$ 0.02& 16.97 $\pm$ 0.04\\        
20/12/2015& 245\,7377.52 &   30.52&                  & 19.12 $\pm$ 0.11& 17.75 $\pm$ 0.02& 17.32 $\pm$ 0.02& 17.02 $\pm$ 0.06\\        
23/12/2015& 245\,7380.43 &   33.43&                  & 19.17 $\pm$ 0.12& 17.87 $\pm$ 0.03& 17.35 $\pm$ 0.03& 16.98 $\pm$ 0.02\\        
19/01/2016& 245\,7407.50 &   60.50&                  &                 & 18.64 $\pm$ 0.03& 18.28 $\pm$ 0.03& 17.86 $\pm$ 0.05\\        
25/01/2016& 245\,7413.27 &   66.27&                  &                 & 18.80 $\pm$ 0.04& 18.30 $\pm$ 0.04& 17.96 $\pm$ 0.04\\        
20/02/2016& 245\,7439.21 &   92.21&                  &                 &                 & 18.98 $\pm$ 0.06& 18.40 $\pm$ 0.06\\        
13/03/2016& 245\,7461.22 &  114.22&                  & 20.39 $\pm$ 0.10& 19.71 $\pm$ 0.04& 19.53 $\pm$ 0.05& 18.93 $\pm$ 0.06\\        
\multicolumn{8}{@{}c}{ \vspace{-2ex}}\\
\multicolumn{8}{@{}c}{\bf ASASSN-16ex \vspace{1ex}}\\
07/05/2016& 245\,7516.38 & $-$6.87& 16.41 $\pm$ 0.02& 17.07 $\pm$ 0.02& 17.03 $\pm$ 0.01& 17.01 $\pm$ 0.01& 17.16 $\pm$ 0.03\\        
12/05/2016& 245\,7521.15 & $-$2.10& 16.22 $\pm$ 0.05& 16.86 $\pm$ 0.01& 16.81 $\pm$ 0.01& 16.79 $\pm$ 0.02& 16.92 $\pm$ 0.04\\        
14/05/2016& 245\,7523.25 &  0.00  & 16.26 $\pm$ 0.03& 16.84 $\pm$ 0.01& 16.78 $\pm$ 0.02& 16.78 $\pm$ 0.01& 16.95 $\pm$ 0.03\\        
16/05/2016& 245\,7525.28 &  2.03  & 16.28 $\pm$ 0.03& 16.85 $\pm$ 0.02& 16.78 $\pm$ 0.01& 16.78 $\pm$ 0.02& 16.95 $\pm$ 0.04\\        
21/05/2016& 245\,7530.25 &  7.00  & 16.59 $\pm$ 0.05& 17.04 $\pm$ 0.03& 16.89 $\pm$ 0.03& 16.90 $\pm$ 0.02& 17.12 $\pm$ 0.04\\        
03/06/2016& 245\,7543.36 &  20.11 &                 & 17.93 $\pm$ 0.01& 17.33 $\pm$ 0.01& 17.21 $\pm$ 0.01& 17.27 $\pm$ 0.02\\        
10/06/2016& 245\,7550.42 &  27.17 &                 & 18.46 $\pm$ 0.01& 17.59 $\pm$ 0.02& 17.30 $\pm$ 0.01& 17.24 $\pm$ 0.02\\        
18/06/2016& 245\,7558.31 &  35.06 & 18.89 $\pm$ 0.05& 19.04 $\pm$ 0.03& 17.95 $\pm$ 0.02& 17.58 $\pm$ 0.04& 17.40 $\pm$ 0.02\\        
21/06/2016& 245\,7561.25 &  38.00 & 19.07 $\pm$ 0.08& 19.14 $\pm$ 0.03& 18.06 $\pm$ 0.02& 17.65 $\pm$ 0.02& 17.45 $\pm$ 0.02\\        
26/06/2016& 245\,7566.37 &  43.12 &                 & 19.33 $\pm$ 0.02& 18.26 $\pm$ 0.01& 17.85 $\pm$ 0.02& 17.55 $\pm$ 0.02\\        
06/07/2016& 245\,7576.17 &  52.92 &                 & 19.47 $\pm$ 0.02& 18.46 $\pm$ 0.01& 18.20 $\pm$ 0.03& 17.94 $\pm$ 0.04\\       
10/07/2016& 245\,7580.16 &  56.91 &                 &                 &                 & 18.36 $\pm$ 0.02&                 \\        
31/07/2016& 245\,7601.13 &  77.88 &                 &                 & 19.04 $\pm$ 0.01& 18.87 $\pm$ 0.02& 18.73 $\pm$ 0.02\\        
14/08/2016& 245\,7615.26 &  92.01 &                 & 20.01 $\pm$ 0.04& 19.28 $\pm$ 0.02& 19.12 $\pm$ 0.02& 18.99 $\pm$ 0.04\\         
17/09/2016& 245\,7649.17 &  125.92&                 & 20.48 $\pm$ 0.10& 19.96 $\pm$ 0.04& 19.88 $\pm$ 0.08& 19.89 $\pm$ 0.12\\        
\hline
\multicolumn{8}{@{}l}{$^a$in days with respect to the epoch of the $B$ band maximum. All magnitudes are in the Vega system.}
\end{tabular}
\label{tab_sn_mag}
\end{table*}

\begin{table*}
\caption{UV--optical photometry of ASASSN-16ex with {\it Swift} UVOT.}
\small
\centering
\begin{tabular}{@{}crcccccc@{}}
\hline
 JD & Phase$^a$ & $uvw2$ & $uvm2$ & $uvw1$ & $u$ & $b$ & $v$\\
\hline
245\,7514.46& $-$8.79& 18.43 $\pm$ 0.18	&18.04 $\pm$ 0.12&17.12 $\pm$ 0.10&	16.28 $\pm$	0.06&17.24 $\pm$ 0.08 &	17.32 $\pm$	0.18 \\
245\,7517.70& $-$5.55& 18.05 $\pm$ 0.15	&17.71 $\pm$ 0.11&17.02 $\pm$ 0.09&	16.14 $\pm$	0.07&17.01 $\pm$ 0.08 &	16.90 $\pm$	0.14 \\
245\,7522.75& $-$0.50& 18.21 $\pm$ 0.17	&18.09 $\pm$ 0.15&17.06 $\pm$ 0.10&	16.10 $\pm$	0.07&16.94 $\pm$ 0.08 &	16.78 $\pm$	0.13 \\
245\,7530.21&    6.96& 18.94 $\pm$ 0.21	&18.71 $\pm$ 0.19&18.05 $\pm$ 0.16&	16.54 $\pm$	0.07&16.98 $\pm$ 0.07 &	16.76 $\pm$	0.11 \\
245\,7532.67&    9.42& 19.85 $\pm$ 0.78	&19.36 $\pm$ 0.67&18.19 $\pm$ 0.26&	16.83 $\pm$	0.14&17.06 $\pm$ 0.12 &	17.02 $\pm$	0.23 \\
245\,7537.52&   14.27& 19.75 $\pm$ 0.50	&19.86 $\pm$ 0.54&18.94 $\pm$ 0.35&	17.27 $\pm$	0.13&17.41 $\pm$ 0.10 &	17.35 $\pm$	0.22 \\
245\,7539.32&   16.07& 20.01 $\pm$ 0.55	&19.91 $\pm$ 0.45&18.92 $\pm$ 0.29&	17.73 $\pm$	0.17&17.50 $\pm$ 0.11 &	17.20 $\pm$	0.16 \\
245\,7541.85&   18.60& 20.28 $\pm$ 1.23	&19.94 $\pm$ 0.51&                &	18.01 $\pm$	0.33&17.60 $\pm$ 0.16 &	17.19 $\pm$	0.27 \\
245\,7544.50&   21.25& 20.16 $\pm$ 0.71	&19.88 $\pm$ 0.50&19.17 $\pm$ 0.44&	18.27 $\pm$	0.33&17.80 $\pm$ 0.14 &	17.40 $\pm$	0.26 \\
245\,7547.84&   24.59&              	&            	 &19.15 $\pm$ 0.24&	18.44 $\pm$	0.15&18.19 $\pm$ 0.14 &	17.74 $\pm$	0.56 \\
245\,7551.21&   27.96&              	&            	 &19.33 $\pm$ 0.18& 18.72 $\pm$	0.12&18.56 $\pm$ 0.13 &	17.66 $\pm$	0.14 \\
245\,7553.54&   30.29&              	&            	 &19.49 $\pm$ 0.31&	18.77 $\pm$	0.20&18.86 $\pm$ 0.24 &	17.79 $\pm$	0.19 \\
245\,7557.45&   34.20&              	&            	 &19.82 $\pm$ 0.49&	18.92 $\pm$	0.29&19.00 $\pm$ 0.30 &	18.02 $\pm$	0.28 \\
245\,7565.88&   42.63&              	&            	 &19.91 $\pm$ 0.43& 19.42 $\pm$	0.35&19.36 $\pm$ 0.29 & 18.28 $\pm$	0.36 \\
245\,7572.28&   49.03&              	&            	 &20.19 $\pm$ 0.35&	19.68 $\pm$	0.25&19.34 $\pm$ 0.18 &	            	 \\
\hline           
\multicolumn{8}{@{}l}{$^a$in days with respect to the epoch of the $B$ band maximum. All magnitudes are in the Vega system.}
\end{tabular}	        	    
\label{tabasn16ex_mag_uvot}	    
\end{table*} 

\subsubsection{UV--optical photometry of ASASSN-16ex using $Swift$ UVOT.}
ASASSN-16ex was also observed with the Ultraviolet/Optical Telescope (UVOT; \citealt{romi05}) onboard the Neil Gehrels {\it Swift} Observatory.  Raw imaging  data in the three broad-band UV filters ($uvw2$: 1928\,\AA, $uvm2$: 2246\,\AA, $uvw1$: 2600\,\AA) and three broad-band optical filters ($u$: 3465\,\AA, $b$: 4392\,\AA, $v$: 5468\,\AA) were accessed from the {\it Swift} database. 
Data processing was done using various packages available in the High Energy Astrophysics Software ({\sc heasoft}\footnote{\url{https://heasarc.gsfc.nasa.gov/docs/software/heasoft/}}), following methods outlined in \cite{pool08}, \cite{brow09, brow14_sousa} and \cite{bree11}. 
The UV--optical magnitudes of ASASSN-16ex are listed in Table~\ref{tabasn16ex_mag_uvot}. 

\subsection{Spectroscopy}
\label{sec_spec_observation}
Medium resolution ($\sim$\,7\,\AA) spectra of SN 2013bz, PSN J0910+5003 and ASASSN-16ex were obtained using the HFOSC-HCT. Two grisms, Gr7 (3500--7000\,\AA) and Gr8 (5200--9200\,\AA), were used to cover the optical wavelength range. A log of spectroscopic observation is  given in Table~\ref{tab_spec_log}. Besides bias and flat frames, spectra of FeNe and FeAr arc lamps were taken for wavelength calibration.  Spectra of spectrophotometric standards were obtained to correct for the instrumental response.
 
Spectroscopic data were reduced using {\sc iraf} in  a standard manner, as discussed in \citet{chak14,chak19}. 
The final spectra were corrected for reddening (refer to Section~\ref{sec_reddening}) and redshift using $z$ = 0.019 for SN 2013bz (source NED), $z$ = 0.034 for PSN J0910+5003 (source NED), and $z$ = 0.04 for ASASSN-16ex \citep{fole18}. 

\begin{table}
\caption{Log of spectroscopic observations of SN 2013bz, PSN J0910+5003 and ASASSN-16ex.}
\small
\centering
\begin{tabular}{@{}lcrc@{}}
\hline
Date & JD$^a$ & Phase$^b$  & Range (\AA)\\
\hline
\multicolumn{4}{@{}c}{\bf SN 2013bz \vspace{1ex}}\\
30/04/2013  & 6413.30    & 3.80 &      3500--7000\\ 
02/05/2013  & 6415.16    & 5.66 &      3500--7000; 5200--9100\\ 
03/05/2013  & 6416.18    & 6.68 &      3500--7000; 5200--9100\\    
06/05/2013  & 6419.26    & 9.76 &      3500--7000\\ 
09/05/2013  & 6422.19    & 12.69&      3500--7000; 5200--9100\\ 
14/05/2013  & 6427.14    & 17.64&      3500--7000; 5200--9100\\ 
20/05/2013  & 6433.25    & 23.75&      3500--7000; 5200--9100\\ 
21/05/2013  & 6434.34    & 24.84&      3500--7000\\ 
27/05/2013  & 6440.22    & 30.72&      3500--7000; 5200--9100\\ 
29/05/2013  & 6442.24    & 32.74&      3500--7000; 5200--9100\\ 
31/05/2013  & 6444.22    & 34.72&      3500--7000; 5200--9100\\     
\multicolumn{4}{@{}c}{ \vspace{-2ex}}\\
\multicolumn{4}{@{}c}{\bf PSN J0910+5003 \vspace{1ex}}\\
17/11/2015&	7344.42&      $-$2.58&3500--7000; 5200--9100\\            
20/11/2015&	7347.44&	 0.44&3500--7000; 5200--9100\\
27/11/2015&	7354.39&	 7.39&3500--7000; 5200--9100\\
28/11/2015&	7355.36& 	 8.36&3500--7000; 5200--9100\\            
09/12/2015&	7366.41&	19.41&3500--7000; 5200--9100\\
20/12/2015&	7377.45&	30.45&3500--7000; 5200--9100\\
19/01/2016&	7407.41&	60.41&3500--7000; 5200--9100\\
22/01/2016&	7410.24&	63.24&3500--7000; 5200--9100\\            
\multicolumn{4}{@{}c}{ \vspace{-2ex}}\\
\multicolumn{4}{@{}c}{\bf ASASSN-16ex \vspace{1ex}}\\
06/05/2016 &  7515.42 & $-$7.83	&3500--7000	\\
08/05/2016 &  7517.41 & $-$5.84 	&3500--7000; 5200--9100	\\
13/05/2016 &  7522.17 & $-$1.08	&3500--7000; 5200--9100	\\
14/05/2016 &  7523.16 & $-$0.09	&3500--7000; 5200--9100	\\
21/05/2016 &  7530.27 &    7.02 	&3500--7000; 5200--9100	\\
11/06/2016 &  7551.18 &   27.93	&3500--7000	\\
19/06/2016 &  7559.35 &   36.10	&3500--7000; 5200--9100	\\
26/06/2016 &  7566.30 &   43.05	&3500--7000	\\
\hline
\multicolumn{4}{@{}l}{$^a$2450000+; $^b$in days from the $B$ band maximum.}
\label{tab_spec_log}
\end{tabular}
\end{table}

\section{Light  curves and Colour Curves}
\label{sec_light_curve}
\begin{figure}
\includegraphics[width=\columnwidth]{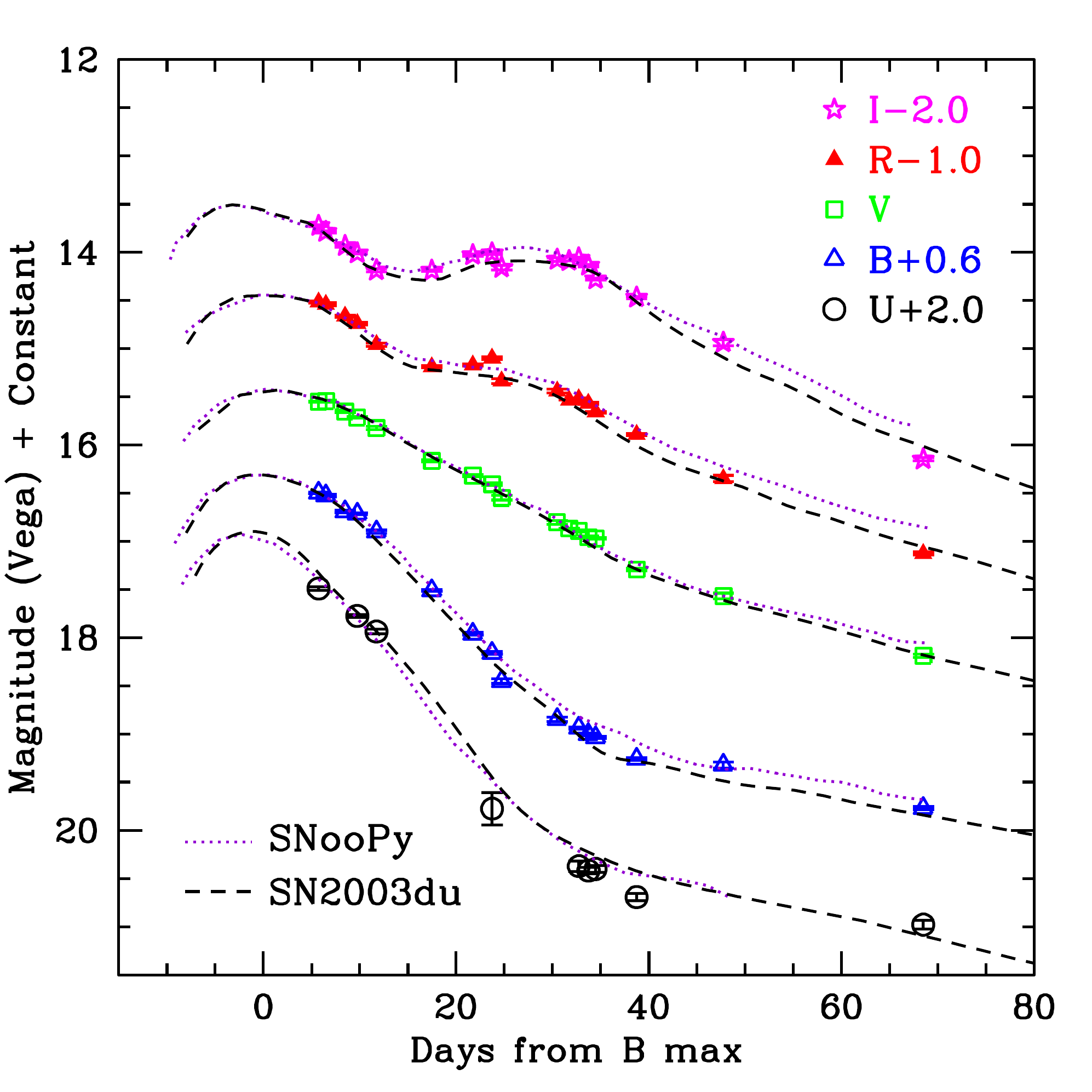}
\caption{$UBVRI$ light curves of SN 2013bz. The light curves are shifted vertically by the
amount indicated in the legend. The phase is measured in days from the $B$ band maximum. The dotted lines represent the template
light curves obtained using the SNooPy code. The dashed lines represent the light curves of the normal Type Ia SN 2003du. }
\label{fig13bz_lc}
\end{figure}

\begin{figure}
\includegraphics[width=\columnwidth]{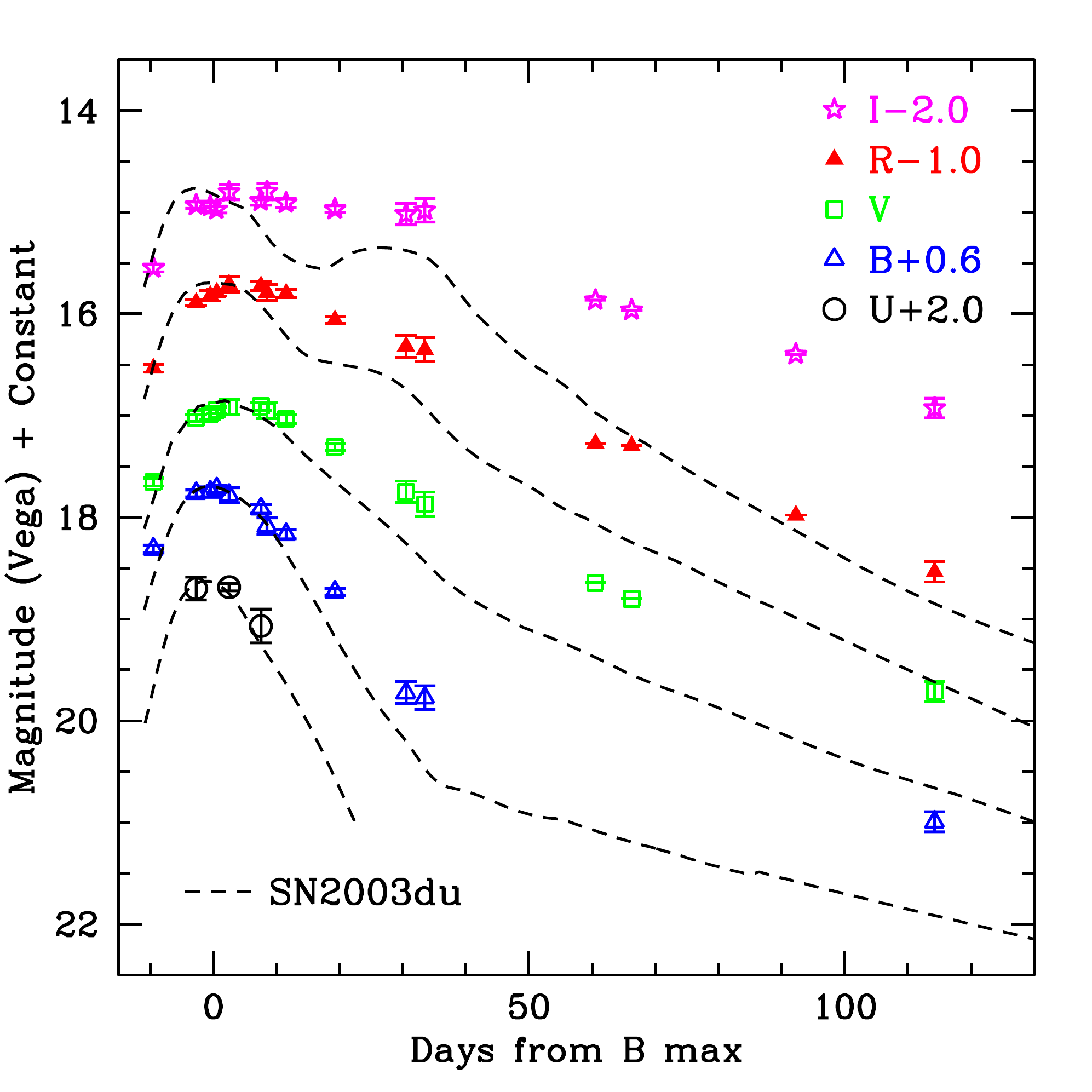}
\caption{$UBVRI$ light curves of PSN J0910+5003. The light curves are shifted vertically by the
amount indicated in the legend. The dashed lines represent the light curves of the normal Type Ia SN 2003du. }
\label{figpsn09_lc}
\end{figure}

\begin{figure}
\includegraphics[width=\columnwidth]{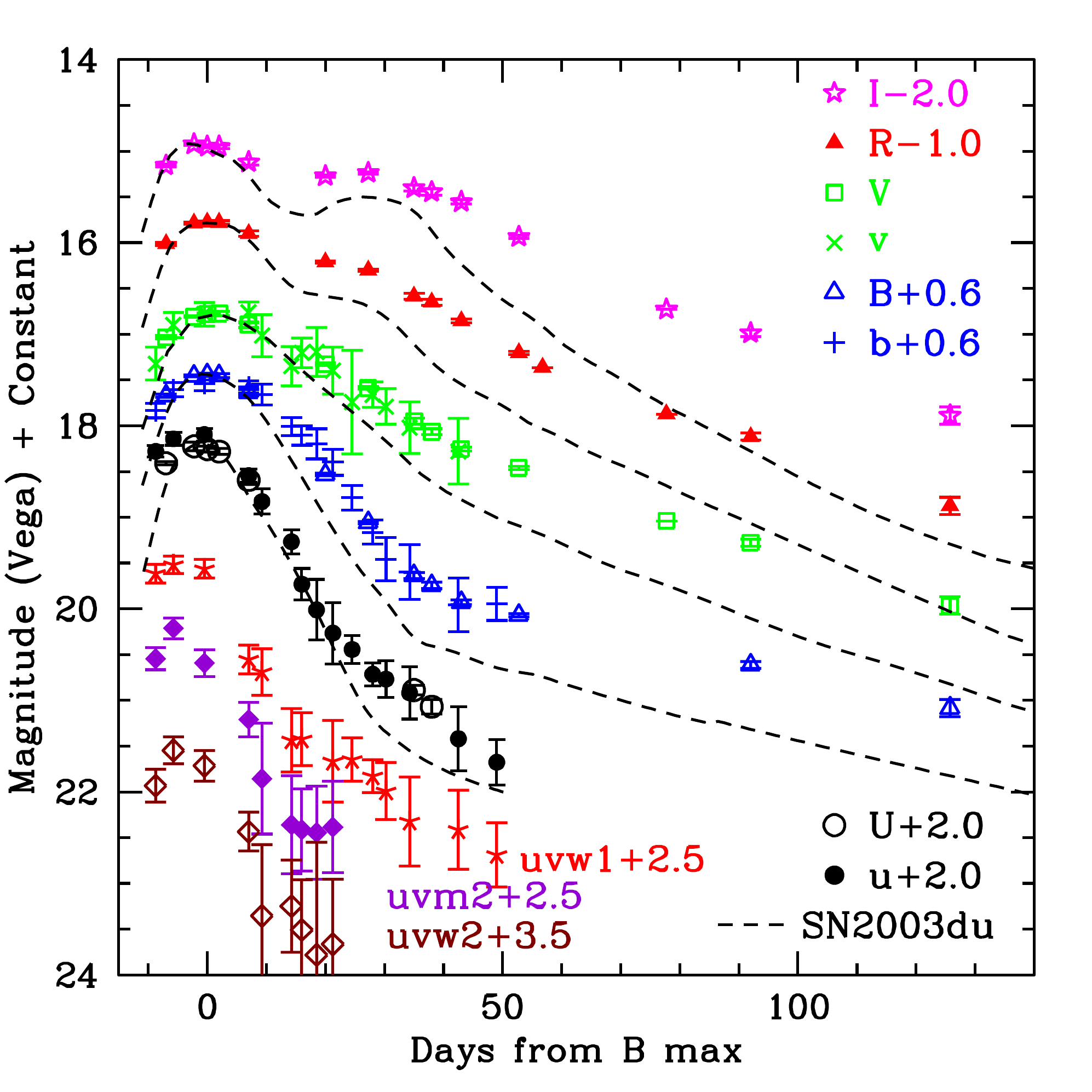}
\caption{$UBVRI$ and {\it Swift} UVOT light curves of ASASSN-16ex. The light curves are shifted vertically by the amount indicated in the legend. The dashed lines represent the light curves of the normal Type Ia SN 2003du. }
\label{figasn16ex_lc}
\end{figure}

The estimated magnitudes of SN 2013bz, PSN J0910+5003 and ASASSN-16ex are plotted in Fig.~\ref{fig13bz_lc}, Fig.~\ref{figpsn09_lc} and Fig.~\ref{figasn16ex_lc}, respectively. UV--optical data of ASASSN-16ex from {\it Swift} UVOT are also plotted in Fig.~\ref{figasn16ex_lc}. 
In order to understand the characteristics of these SNe; we derived various photometric parameters from their light curves and are  listed in  Table~\ref{tab_peak} and  Table~\ref{tab_lc_parameter}.

Our observation of SN 2013bz started in the post-maximum phase;  hence we used max-model of SNooPy \citep{burn11}  package to fit templates and derive  light curve parameters.  The best-matching template light curves are over-plotted onto the observed light curves (Fig.~\ref{fig13bz_lc}).  SN 2013bz reached maximum brightness in the $B$ band on JD 245\,6409.5 $\pm$ 0.8 with an apparent brightness of 15.71 $\pm$ 0.04\,mag. 
Similar to the normal and 91T--like  SNe Ia, in  SN 2013bz, the maxima in $U$ and $I$ bands precede, and those in the $V$ and $R$ bands follow the $B$  band maximum (refer to Table~\ref{tab_peak}). 
The decline in the $B$ band brightness 15 days after the maximum is estimated as $\Delta m_{15}(B)$ = 0.91 $\pm$ 0.04\,mag and $\Delta m_{15}(B)_{true}$ =  0.92 $\pm$ 0.04 \citep{phil99,fola10}. The estimated decline rate parameter for SN 2013bz is smaller than the normal SNe Ia, suggesting it to be a luminous event \citep{phil99}. The decline rates in other bands are listed in Table~\ref{tab_lc_parameter}.  A distinct and pronounced secondary peak in the $I$ band light curve, a characteristic feature of normal and 91T--like SNe Ia, is seen  $\sim$ +30\,d. 
A small hump is also seen in the $R$ band light curve at a similar phase. The $I$ band secondary maximum timing
agrees with its derived $\Delta m_{15}(B)$ \citep{hamu96_temp,kase06,fola10}.  The light curves of a typical Ia event SN 2003du \citep*{anup05} are also plotted along with SN 2013bz in Fig.~\ref{fig13bz_lc}. From a quick comparison, it is apparent  that the light curves of SN 2013bz are similar to a normal Ia event. 

In the case of PSN J0910+5003 and ASASSN-16ex, our observations started well before the maximum phase. We fitted  a cubic spline function to the data points around the maximum to estimate the date of maximum and magnitude at maximum. 
PSN J0910+5003 reached maximum brightness in the $B$ band on JD 245\,7347.0 $\pm$ 0.6 with 17.13 $\pm$ 0.04\,mag. The decline rate parameter is estimated as $\Delta m_{15}(B)$ = 0.69 $\pm$ 0.04 and $\Delta m_{15}(B)_\text{true}$ = 0.70 $\pm$ 0.04.
The decline rate of PSN J0910+5003 is slower than the normal SNe Ia. 
The important photometric parameters derived for PSN J0910+5003 are listed in Table~\ref{tab_lc_parameter}. 

ASASSN-16ex reached maximum light on JD 245\,7523.25 $\pm$ 0.5 with 16.84 $\pm$ 0.03\,mag. 
The decline rate parameter for ASASSN-16ex is estimated as $\Delta m_{15}(B)$ = 0.72 $\pm$ 0.03 and  $\Delta m_{15}(B)_\text{true}$ = 0.73 $\pm$ 0.03.  The important light curve parameters derived for ASASSN-16ex (including data from UVOT) are listed in Table~\ref{tab_lc_parameter}.

The slow declining nature of PSN J0910+5003 and  ASASSN-16ex is obvious from an immediate comparison of their light curves (refer to   Fig.~\ref{figpsn09_lc}, \ref{figasn16ex_lc}) with the light curves of the normal Type Ia SN 2003du.  It is interesting to note that for both PSN J0910+5003 and ASASSN-16ex, $\Delta m_{15}(B)$ is close to 09dc--like SNe Ia: 
SN 2009dc ($\Delta m_{15}(B)$ = 0.71; \citealt{taub11}), 
SN 2007if ($\Delta m_{15}(B)$ = 0.71; \citealt{scal10}), 
SN 2006gz ($\Delta m_{15}(B)$ = 0.69; \citealt{hick07}),
ASASSN-15pz ($\Delta m_{15}(B)$ = 0.67; \citealt{chen19}, and
ASASSN-15hy ($\Delta m_{15}(B)$ = 0.72; \citealt{luj21}). 

Timing of maximum brightness in $U$, $V$, $R$ and $I$ bands for PSN J0910+5003, ASASSN-16ex and some other well-studied SNe Ia are listed in Table~\ref{tab_peak}. In PSN J0910+5003, the $I$ band peak appears much later (6\,d) than the $B$ band, similar to 09dc--like SNe Ia. In contrast, the $I$ band peak in ASASSN-16ex follows the trend of normal SNe Ia.  

\begin{figure}
\includegraphics[width=\columnwidth]{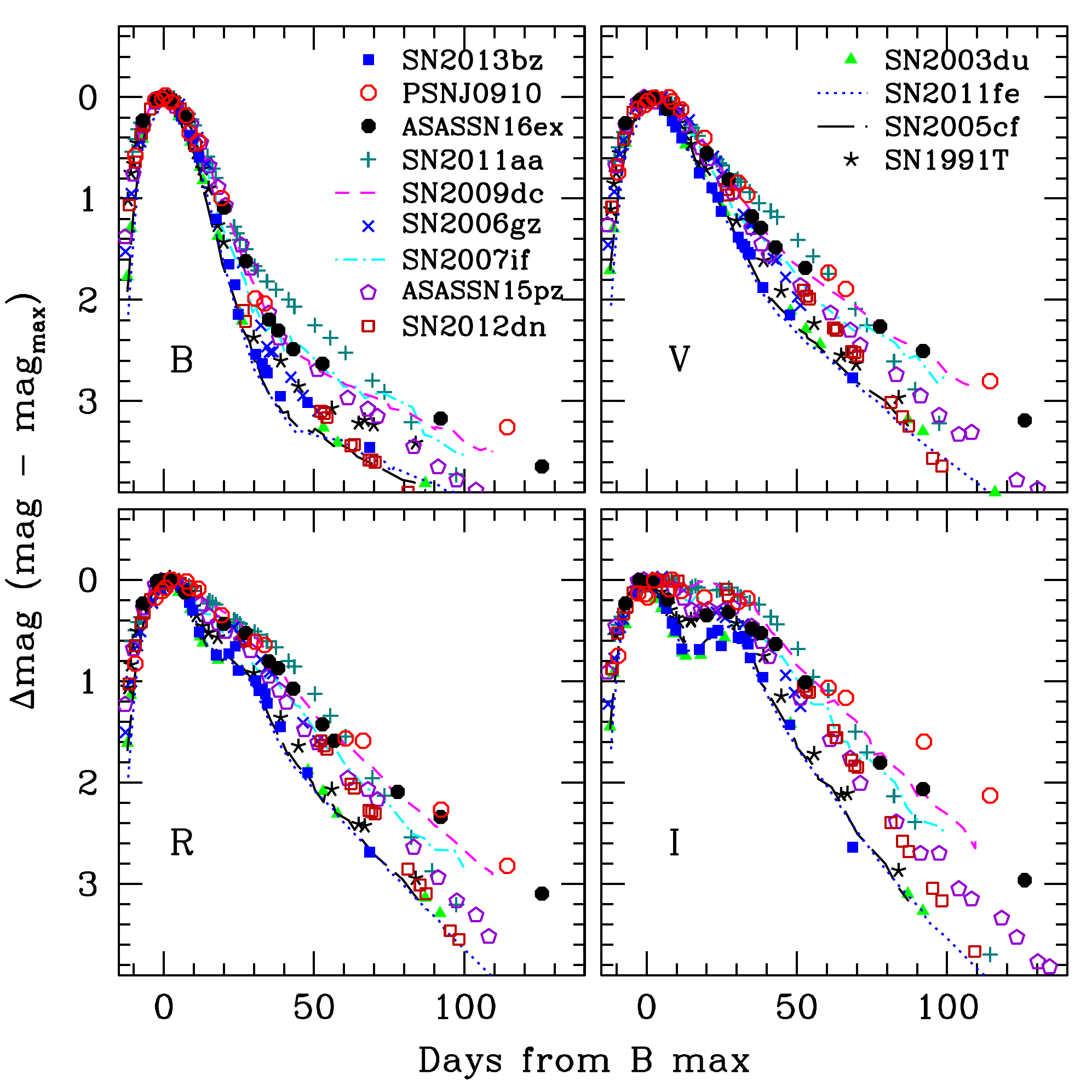}
\caption{$BVRI$ light curves of SN 2013bz, PSN J0910+5003 and ASASSN-16ex are compared with those of some other well-studied SNe Ia. The light curves are shifted to match their peak brightness and the epoch of the $B$ band maximum.}
\label{fig_lc_comp}
\end{figure}

In Fig.~\ref{fig_lc_comp},  $BVRI$ light curves of SN 2013bz, PSN J0910+5003 and ASASSN-16ex are compared with those of well-studied normal  SNe Ia: 
SN 2003du ($\Delta m_{15}(B)$ = 1.04; \citealt{anup05}),
SN 2005cf ($\Delta m_{15}(B)$ = 1.11; \citealt{past07}),
SN 2011fe ($\Delta m_{15}(B)$ = 1.07; \citealt{rich12,vink12,brow12}),
the luminous SN 1991T ($\Delta m_{15}(B)$ = 0.95; \citealt{lira98}), 
and 09dc--like SNe Ia: 
SN 2006gz (\citealt{hick07}), 
SN 2007if (\citealt{scal10}), 
SN 2009dc (\citealt{taub11}), 
ASASSN-15pz (\citealt{chen19}),
SN 2012dn ($\Delta m_{15}(B)$ = 0.92; \citealt{chak14}) and
SN 2011aa ($\Delta m_{15}(B)$ = 0.59; \citealt{dutt22}). 
All the light curves are shifted to match their peak brightness and the epoch of the $B$ band maximum.

\begin{figure}
\includegraphics[width=\columnwidth]{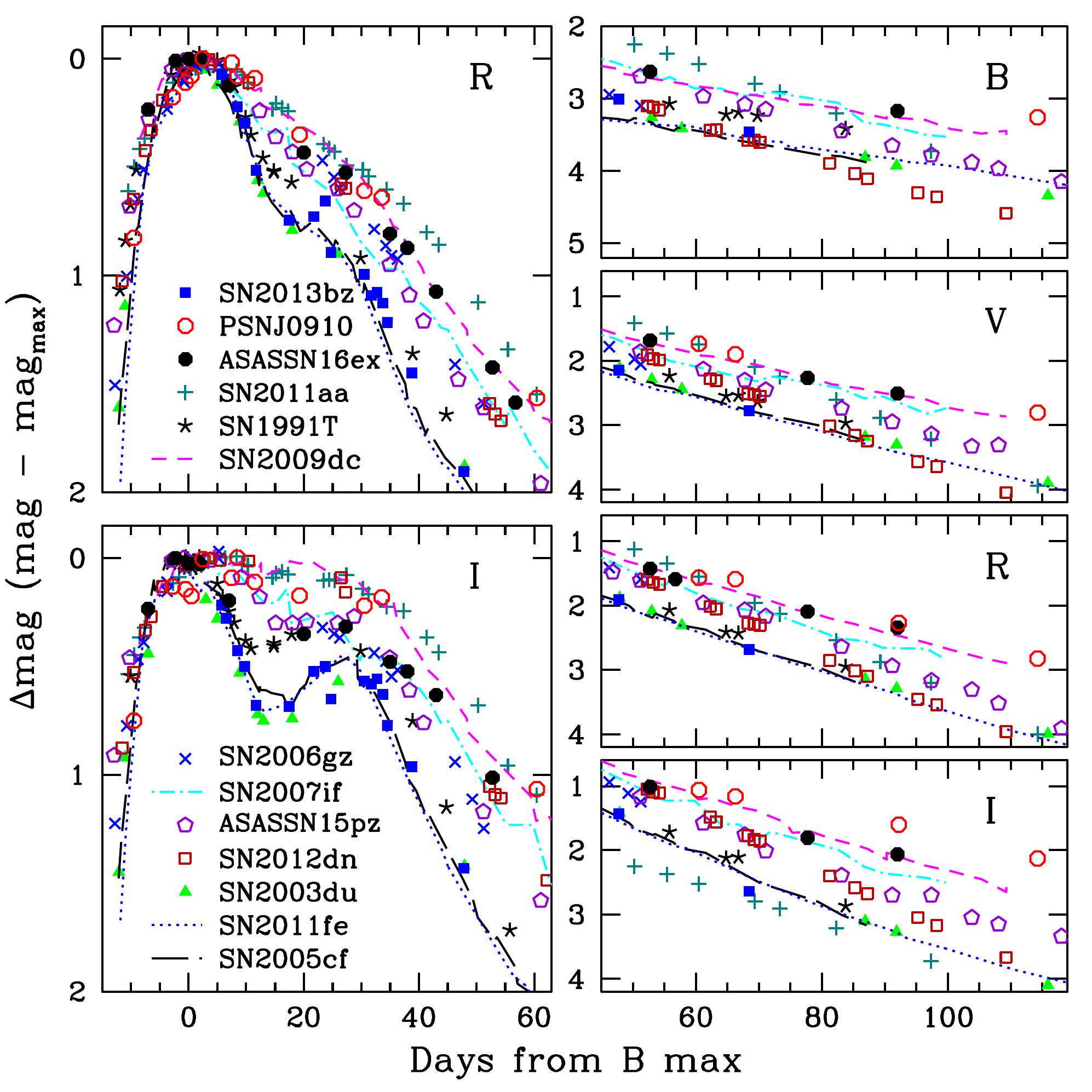}
\caption{A zoomed view of the light curve comparison. {\it Left}: $R$ and $I$ band light curves during secondary maximum. {\it Right}: $BVRI$ light curves during the late phase. }
\label{fig_lc_comp_zoomed}
\end{figure}

The comparison shows that SN 2013bz closely follows the light curves of normal type Ia SNe 2003du/05cf/11fe in  $BVRI$ bands. The trough in between $I$ band primary and  secondary maxima, the strength of the $I$ band secondary maximum, and the shoulder in  the $R$ band also match those of normal SNe Ia (refer to Fig.~\ref{fig_lc_comp_zoomed} for a zoomed view). 
Though the decline rate parameter, $\Delta m_{15}(B)$, is lower than that of a typical SN Ia, SN 2013bz appears to be a perfectly normal type Ia event. 

\begin{figure}
\includegraphics[width=\columnwidth]{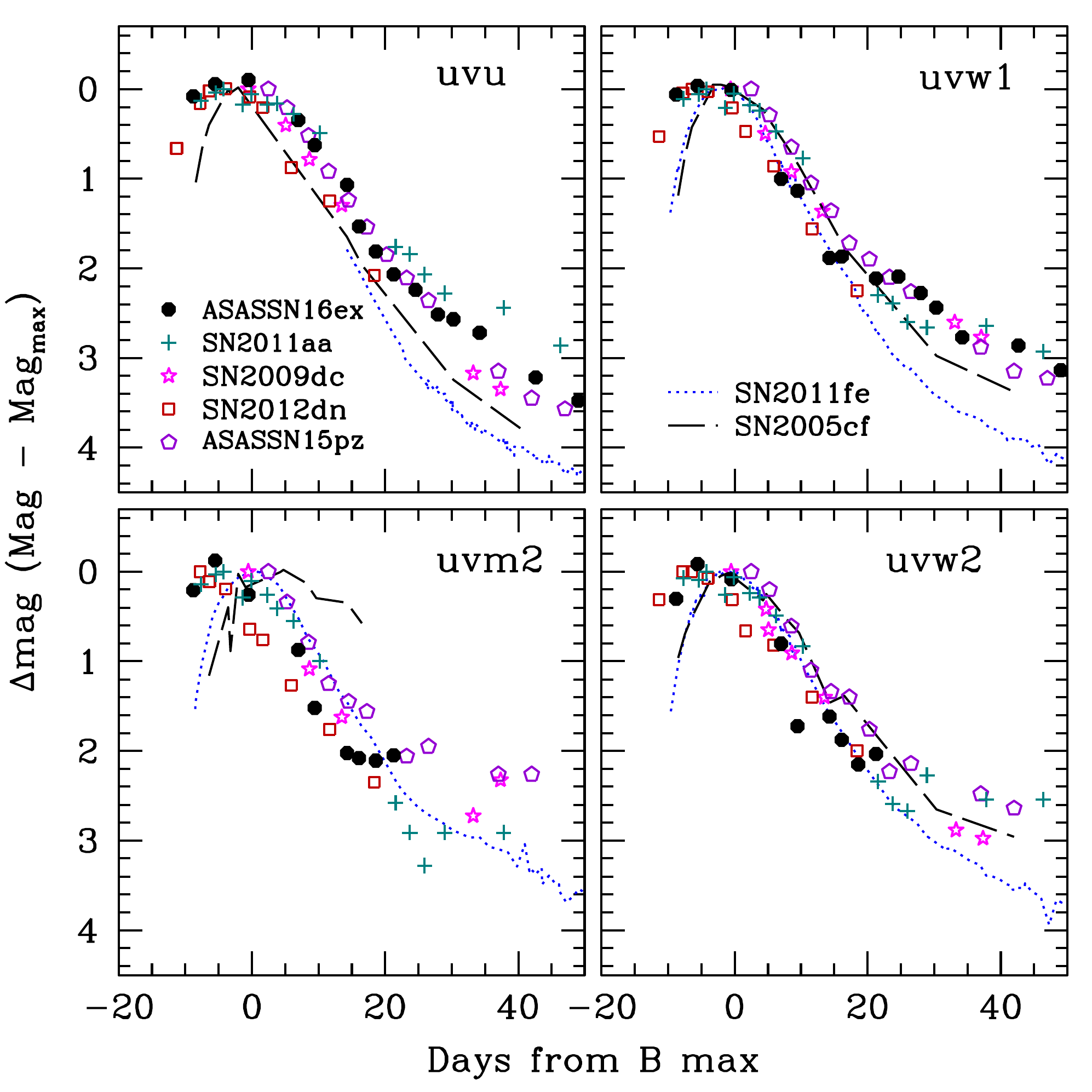}
\caption{$UV$ light curves of ASASSN-16ex are compared with some well-studied SNe Ia. The light
curves are shifted to match their peak brightness and the epoch of the $B$ band maximum.}
\label{fig_lc_comp_uv}
\end{figure}

On the other hand,  $BVRI$ light curves of PSN J0910+5003 and ASASSN-16ex are broad and  closely follow the 09dc--like events. 
Similar to 09dc--like SNe,  $I$ band light curves of PSN J0910+5003 and ASASSN-16ex appear flat during the early post-maximum phase (till $\sim$ 40\,d); in the later phase they also follow 09dc--like SNe. An enhanced fading is seen in the light curves of 09dc--like SNe in the nebular phase. In some events, e.g., SN 2012dn, ASASSN-15pz and LSQ14fmg, it was observed quite early (1--3 months from the maximum phase). 
This is not observed in PSN J0910+5003 and ASASSN-16ex until our last observations ($\sim$ 126\,d, refer to Fig.~\ref{fig_lc_comp_zoomed}, right panel).

The decline rates of ASASSN-16ex and PSN J0910+5003 in $BVRI$ bands during 50--100\,d are estimated by a linear fit and listed in Table~\ref{tab_lc_parameter}. The late phase decline of ASASSN-16ex and PSN J0910+5003 light curves are comparable to SNe 2007if/09dc. 

\begin{table}
\caption{Timing of maximum light in $UVRI$ bands (in days) with respect to the $B$ band maximum.}
\small
\centering
\begin{tabular}{@{}lccccl@{}}
\hline
Object & U & V  & R & I & Ref.$^a$\\
\hline
SN 2013bz      & $-$2.0 & +0.2 & +1.0   & $-$3.0   & 1      \\
PSN J0910+5003 &    -   & +5.0 & +4.0 & +6.0   & 1     \\
ASASSN-16ex    & $-$1.5 & +1.5 &   0  & $-$1.8 & 1      \\
SN 2006gz      & $-$3.0 & +2.0 &   +6.0 &  +6.0 & 2 \\
SN 2009dc      & $-$2.0 & 0    & +1.0   &   +2   & 3 \\
SN 2012dn      & $-$2.5 & +1.6 & +2.6 &   +3   & 4 \\
ASASSN-15pz    &    -   &$\sim$0& +1 &  +0.4  & 5 \\
ASASSN-15hy    & $-$6.0 & +3.0 &  +5.5  & +7.3  & 6 \\
LSQ14fmg       &  -     &$-$1.0 &  +1  &  +1& 7 \\
SN 2020esm     & $-$4.0 &$\sim$0& +3.5 & +5 & 8  \\
SN 2020hvf    & $-$4.0 & +1.0 & +1.8 & +2.7 & 9 \\
SN 2011aa      & $-$0.3 & +3.8 & +4.5 & +7.1 & 10 \\
SN 2003du      & $-$1.3 & +1.0 & +0.3 & $-$1.9 & 11 \\
SN 2005cf      & $-$1.6 & +1.3 & +0.6 & $-$2.0 & 12 \\
SN 2011fe      & $-$1.4 & +1.0 & +0.6 & $-$2.9 & 13, 14 \\
SN 1991T       & $-$1.7 & +2.6 & +1.4 & $-$0.4 & 15 \\
\hline
\multicolumn{6}{@{}l}{$^a$1. This work, 2. \citet{hick07} } \\
\multicolumn{6}{@{}l}{3. \citet{taub11} } \\
\multicolumn{6}{@{}l}{4. \citet{chak14}, 5. \citet{chen19} }\\
\multicolumn{6}{@{}l}{6. \citet{luj21}, 7. \citet{hsia20} }\\
\multicolumn{6}{@{}l}{8. \citet{dimi22}, 9. \citet{jian21} }\\
\multicolumn{6}{@{}l}{10. \citet{dutt22}, 11. \citet{anup05}}\\
\multicolumn{6}{@{}l}{12. \citet{past07}, 13. \citet{rich12}  }\\
\multicolumn{6}{@{}l}{14. \citet{pere13}, 15. \citet{lira98} }
\label{tab_peak}
\end{tabular}
\end{table}

\begin{figure}
\includegraphics[width=\columnwidth]{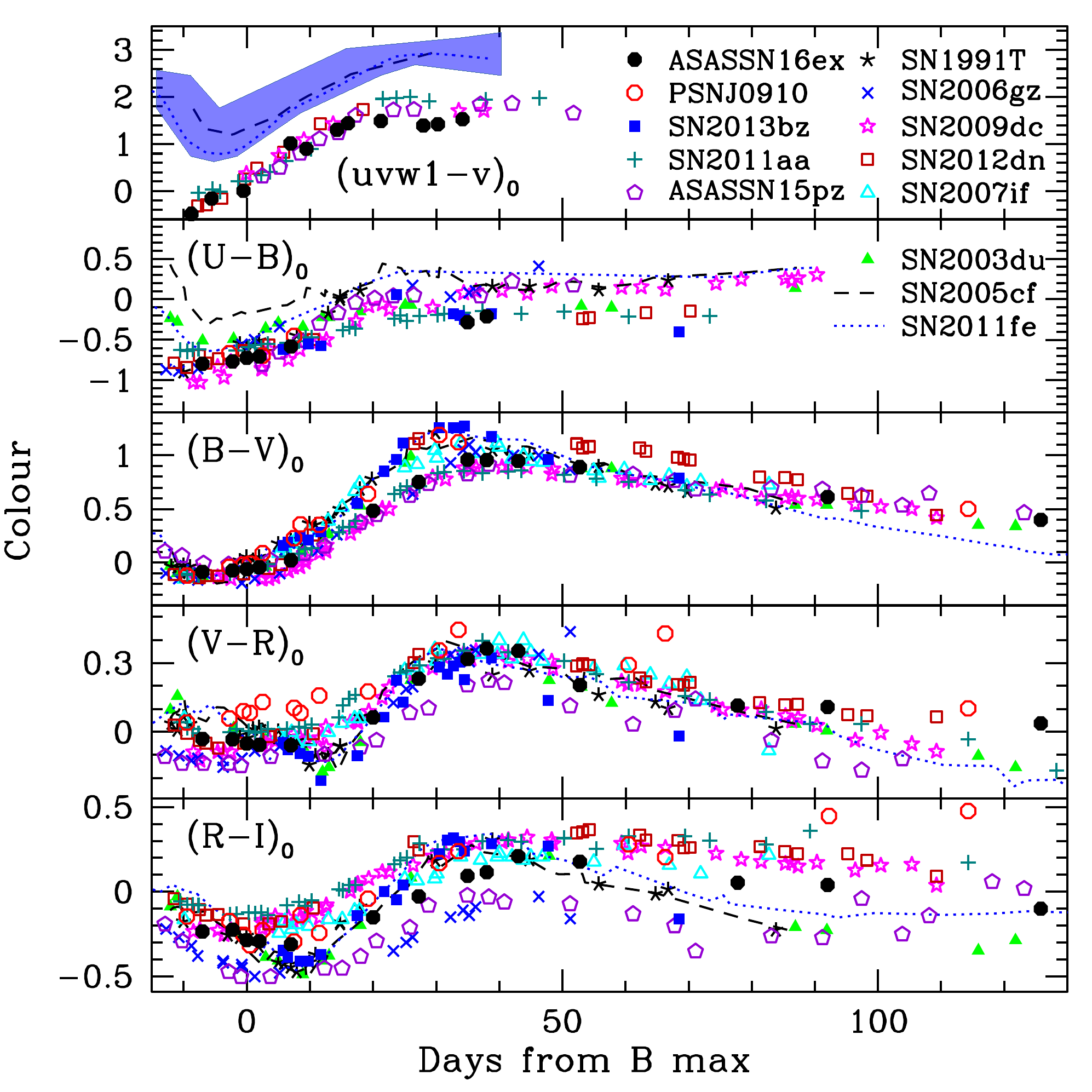}
\caption{The $(uvw1-v)$, $(U-B)$, $(B-V)$, $(V-R)$ and $(R-I)$ colour curves of SN 2013bz, PSN J0910+5003 and ASASSN-16ex are compared with those of other well-studied SNe Ia. The shaded region in the top panel shows the ($uvw1 - v$) colour evolution of normal (NUV-blue + NUV-red, \citealt{miln13,brow14}) SNe Ia.}
\label{fig_colour_comp}
\end{figure}

In Fig.~\ref{fig_lc_comp_uv}, the $UV$ light curves of ASASSN-16ex in $uvu$, $uvw2$, $uvm2$, and $uvw1$ bands are compared with normal and 09dc--like SNe Ia.
It is clear that the light curves of ASASSN-16ex in all the $UV$ bands are distinct from  normal SNe 2005cf/11fe and closely follow the 09dc--like SNe.  Similar to 09dc--like SNe, the $UV$ light curves of ASASSN-16ex in $uvw2$, $uvm2$, and $uvw1$ reached maximum light a few days  before the normal SNe 2005cf/11fe. Also, the $UV$ light curves of ASASSN-16ex and 09dc--like SNe are broader than the normal events. 

The reddening (refer to Section~\ref{sec_reddening}) corrected $(uvw1-v)$, $(U-B)$, $(B-V)$, $(V-R)$ and $(R-I)$ colour curves of SN 2013bz, PSN J0910+5003 and ASASSN-16ex are plotted in Fig.~\ref{fig_colour_comp}. 
Colour curves of well-studied SNe Ia (used in the light curve comparison) are also plotted in Fig.~\ref{fig_colour_comp}. They are corrected for  reddening, as mentioned in their respective references.

The $(uvw1-v)$ colour evolution of ASASSN-16ex is similar to 09dc--like SNe. 
The normal SNe 2005cf/11fe are redder in the early phase and evolve to blue with the bluest value a few days before the $B$ band maximum. Their $(uvw1-v)$ colour evolution makes a characteristic `V'-shape pattern \citep{miln13,brow14}. However, ASASSN-16ex is bluer by $\sim$1\,mag  in the early phase and reddens monotonically, similar to other 09dc--like SNe. 
\citet{miln13} have divided normal SNe Ia into NUV-blue and NUV-red categories based on the (near-UV $-$ optical) colour evolution. 
The shaded region in Fig.~\ref{fig_colour_comp} ({\it top panel}) shows the colour evolution of normal SNe Ia (which includes  NUV-blue and NUV-red objects; \citealt{miln13,brow14}). With a significantly blue (near-UV $-$ optical) colour, ASASSN-16ex and 09dc--like SNe Ia are clearly distinguishable  from normal SNe Ia. This colour criterion can  be used to separate  09dc--like objects from the SNe Ia sample with a precise host extinction estimate.
The $(U-B)$ colour of SN 2013bz, PSN J0910+5003 and ASASSN-16ex is bluer relative to normal events. 

The $(B-V)$ colour of SN 2013bz follows those of normal SNe Ia, while ASASSN-16ex follows the evolution of 09dc--like SNe 2006gz/09dc/ASASSN-15pz in the complete phase range.   
The 09dc--like SNe 2006gz/09dc/ASASSN-15pz have a bluer curve during 0 to +50\,d, while normal SNe 2003du/05cf/11fe and 91T have a relatively red colour. PSN J0910+5003 appears to follow the redder path similar to normal SNe Ia. 

The $(V-R)$ colour of ASASSN-16ex is similar to SN 2009dc up to $\sim$ 100\,d; PSN J0910+5003 have a more gradual and redder colour evolution. The $(R-I)$ colour of ASASSN-16ex is a little bluer but similar to SN 2009dc, while PSN J0910+5003 have a colour similar to SN 2007if.
Compared to normal events, the $(V-R)$ and $(R-I)$ colours of 09dc--like SNe show less evolution, i.e., red to blue change, because of their broad $RI$ light curves. The scatter in the colours within the 09dc--like SNe group can be attributed to their differing $RI$ light curves. The $(V-R)$ and $(R-I)$ colours of SN 2013bz are similar to normal events. 

\begin{table*}
\centering
\caption{Photometric parameters of SN 2013bz, PSN J0910+5003 and ASASSN-16ex.}
\small
\begin{tabular}{@{}lccccc@{}}
\hline
Band & JD (max)  & $m_{\lambda}^\text{max}$ &$M_{\lambda}^{\text{max}}$&  $\Delta m_{15}(\lambda)$ & Decline rate$^a$ \\  
\hline
  &\multicolumn{4}{@{}c}{\bf SN 2013bz}                                & (35--70\,d) \vspace{1ex}\\
$U$ &245\,6407.2 $\pm$ 0.8 &14.93 $\pm$ 0.06 &$-$20.53 $\pm$ 0.20 &1.21 $\pm$ 0.06 & 1.52 \\
$B$ &245\,6409.5 $\pm$ 0.8 &15.71 $\pm$ 0.04 &$-$19.61 $\pm$ 0.20 &0.91 $\pm$ 0.04 & 2.07 \\
$V$ &245\,6409.7 $\pm$ 0.8 &15.42 $\pm$ 0.04 &$-$19.70 $\pm$ 0.20 &0.57 $\pm$ 0.04 & 3.48 \\
$R$ &245\,6410.2 $\pm$ 0.8 &15.44 $\pm$ 0.07 &$-$19.53 $\pm$ 0.20 &0.66 $\pm$ 0.07 & 4.38 \\
$I$ &245\,6406.3 $\pm$ 0.8 &15.51 $\pm$ 0.09 &$-$19.30 $\pm$ 0.20 &0.59 $\pm$ 0.09 & 5.62 \\ 
\multicolumn{6}{@{}c}{ \vspace{-2ex}}\\
     &\multicolumn{4}{@{}c}{\bf PSN J0910+5003}                          &  (50--100\,d) \vspace{1ex}  \\
$B$ &245\,7347.0 $\pm$ 0.6 &17.13 $\pm$ 0.05&$-$19.44 $\pm$ 0.20 &0.69 $\pm$ 0.05 & 1.51 \\
$V$ &245\,7352.0 $\pm$ 0.8 &16.90 $\pm$ 0.04&$-$19.48 $\pm$ 0.20 &0.44 $\pm$ 0.04 & 1.96 \\
$R$ &245\,7351.2 $\pm$ 0.8 &16.70 $\pm$ 0.05&$-$19.54 $\pm$ 0.20 &0.35 $\pm$ 0.04 & 2.43 \\
$I$ &245\,7353.0 $\pm$ 0.9 &16.80 $\pm$ 0.05&$-$19.29 $\pm$ 0.20 &0.18 $\pm$ 0.05 & 1.95 \\
\multicolumn{6}{@{}c}{ \vspace{-2ex}}\\
       & \multicolumn{4}{@{}c}{\bf ASASSN-16ex}                                         & (50--100\,d) \vspace{1ex}  \\     
$uvw2$ &245\,7517.7 $\pm$ 0.6 &18.05 $\pm$ 0.15 & $-$18.80 $\pm$ 0.30 & 1.27 $\pm$ 0.15& - \\
$uvm2$ &245\,7517.7 $\pm$ 0.6 &17.71 $\pm$ 0.11 & $-$19.36 $\pm$ 0.30 & 1.51 $\pm$ 0.11& - \\
$uvw1$ &245\,7517.7 $\pm$ 0.6 &17.02 $\pm$ 0.09 & $-$19.75 $\pm$ 0.30 & 1.26 $\pm$ 0.09& - \\
$U$    &245\,7521.7 $\pm$ 0.7 &16.23 $\pm$ 0.05 & $-$20.47 $\pm$ 0.20 & 0.91 $\pm$ 0.05& - \\
$B$    &245\,7523.2 $\pm$ 0.5 &16.84 $\pm$ 0.03 & $-$19.78 $\pm$ 0.20 & 0.72 $\pm$ 0.03& 2.14\\
$V$    &245\,7524.7 $\pm$ 0.5 &16.77 $\pm$ 0.03 & $-$19.71 $\pm$ 0.20 & 0.43 $\pm$ 0.03& 2.14 \\
$R$    &245\,7523.2 $\pm$ 0.6 &16.78 $\pm$ 0.03 & $-$19.62 $\pm$ 0.20 & 0.35 $\pm$ 0.03& 2.54 \\
$I$    &245\,7521.5 $\pm$ 0.6 &16.94 $\pm$ 0.04 & $-$19.35 $\pm$ 0.20 & 0.31 $\pm$ 0.04& 2.97\\
\hline
\multicolumn{6}{@{}l}{$^a$in unit of mag\,(100\,d)$^{-1}$.}
\end{tabular}
\label{tab_lc_parameter}
\end{table*}

\section{ABSOLUTE LIGHT CURVES and BOLOMETRIC LUMINOSITY}
\label{sec_abs_bol_luminosity}
\subsection{Reddening Estimate}
\label{sec_reddening}

The Galactic reddening for SN 2013bz is $E(B-V)_\text{Gal}$ = 0.04 $\pm$ 0.002\,mag \citep{schl11}.  
Empirical relations \citep{phil99,alta04,rein05,wang06,fola10} suggest a large total reddening, $E(B-V)_\text{total}$ = 0.30 -- 0.40\,mag.
SNooPy fit suggests a reddening of $E(B-V)_\text{total}$ = 0.21 $\pm$ 0.06\,mag.
A strong Na\,{\sc i}\,D feature is seen at the rest frame of the host galaxy in the spectra of SN 2013bz with an average equivalent width (EW) of 1.9 $\pm$ 0.2\,\AA.
The Milky way component of Na\,{\sc i}\,D feature is not detectable. The measured EW (Na\,{\sc i}\,D) translates into $E(B-V)_\text{host}$ = 0.29 $\pm$ 0.03\,mag \citep*{tura03}, close to the value obtained using  other empirical relations. 
We found that the observed $(B-V)$ colour curve of SN 2013bz matches well with normal SNe Ia (refer to Fig.~\ref{fig_colour_comp}) when it is dereddened by $E(B-V)_\text{total}$ = 0.19 $\pm$ 0.04\,mag. This estimate is close to the value suggested by the SNooPy fit and is used for further analysis.

The Galactic reddening for PSN J0910+5003 is $E(B-V)_\text{Gal}$ = 0.02 $\pm$ 0.001\,mag \citep{schl11}. The empirical relations using photometry  suggest $E(B-V)_\text{total} >$ 0.25\,mag.
The Na\,{\sc i}\,D features are not detected in the spectra of PSN J0910+5003. The observed $(B-V)$ colour curve of PSN J0910+5003 matches well with other well-studied SNe Ia (refer to Fig.~\ref{fig_colour_comp}) after correcting it  by $E(B-V)_\text{total}$ = 0.18 $\pm$ 0.05\,mag, which we use in our analysis. 

The Galactic reddening for ASASSN-16ex is $E(B-V)_\text{Gal}$ = 0.03 $\pm$ 0.001\,mag \citep{schl11}. The empirical relations suggest $E(B-V)_\text{total} \sim$  0.11 -- 0.18\,mag. No feature of Na\,{\sc i}\,D is detected in the spectra of ASASSN-16ex. It was shown (refer to Section~\ref{sec_light_curve}) that the photometric properties of ASASSN-16ex closely resemble 09dc--like SNe Ia. \citet{chen19} found that among 09dc--like SNe, ASASSN-15pz suffered minimal reddening, and its colour evolution can be used to estimate the reddening of similar objects. The observed $(B-V)$ colour curve of ASASSN-16ex matches well with ASASSN-15pz and other 09dc--like objects after a correction by $E(B-V)_\text{total}$ = 0.12 $\pm$ 0.04\,mag. Hence, we use this value of total reddening in further analysis.

\begin{figure}
\includegraphics[width=\columnwidth]{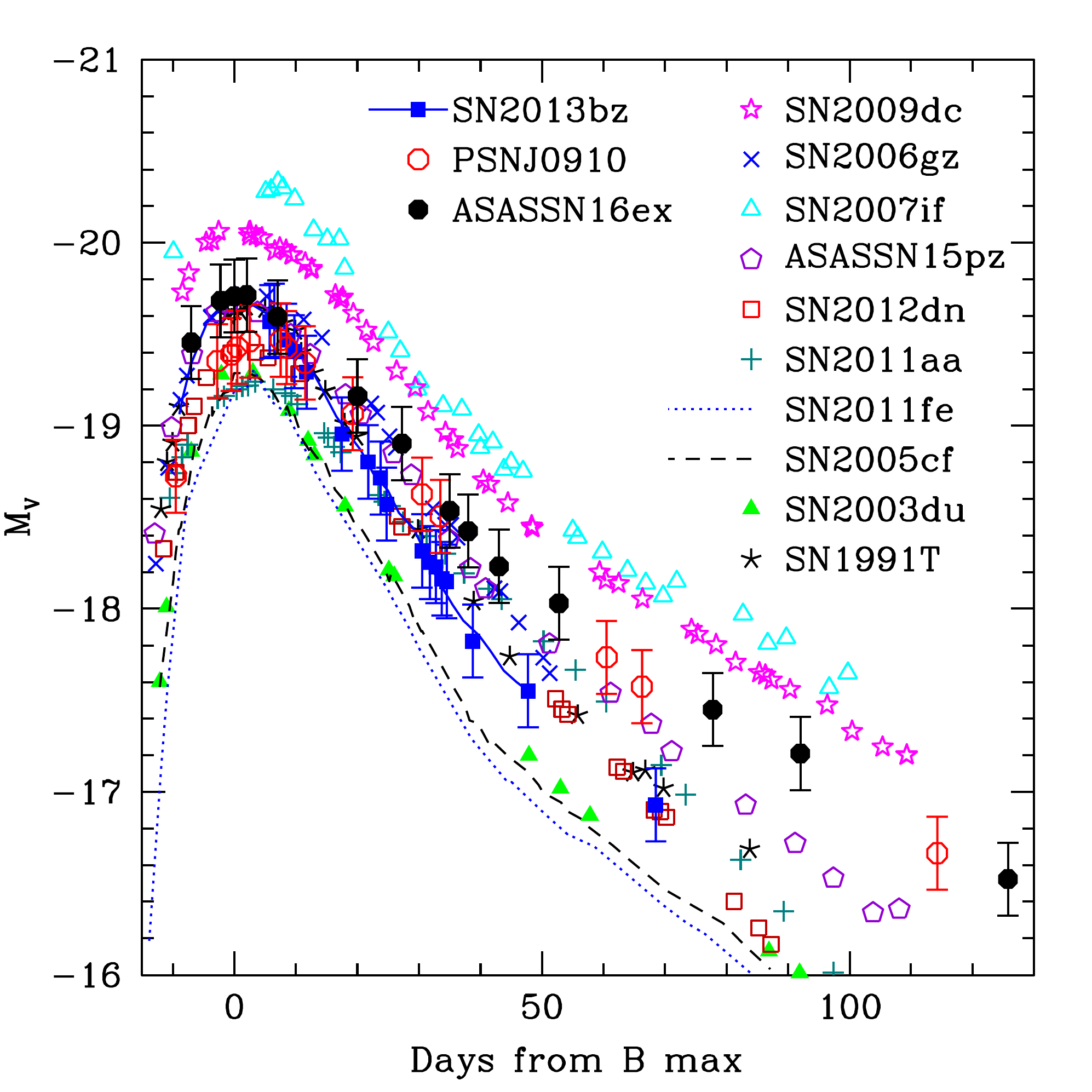}
\caption{Absolute $V$ band light curves of SN 2013bz, PSN J0910+5003
and ASASSN-16ex are compared with those of other well-studied SNe Ia.}
\label{fig_Vabs_comp}
\end{figure}

\subsection{Absolute Magnitudes}
\label{sec_abs_mag}
Recession velocity corrected for infall of the Local Group towards Virgo cluster is 5\,807 $\pm$ 45\,km\,sec$^{-1}$ for PGC 170248, and for UGC 4812, it is 10\,503 $\pm$ 31\,km\,sec$^{-1}$ (\citealt{moul00}; source NED). With $H_{0}$ = 72 $\pm$ 5\,km\,sec$^{-1}$\,Mpc$^{-1}$ \citep{free01}, distances 
of SN 2013bz and PSN J0910+5003 are calculated as 80.65 $\pm$ 5.6\,Mpc ($\mu$ = 34.53 $\pm$ 0.20\,mag) and 145.88 $\pm$ 10\,Mpc ($\mu$ = 35.82 $\pm$ 0.20\,mag), respectively. For SDSS J171023.63+262350.3 (host of ASASSN-16ex), using $z$ = 0.04 \citep{fole18}, we derived a distance of 166.67 $\pm$ 12\,Mpc ($\mu$ = 36.11 $\pm$ 0.20\,mag).  

Using the estimated distances, reddening values and \citet*{card89} extinction law with  $R_V = 3.1$, the peak absolute magnitudes of SN 2013bz, PSN J0910+5003 and ASASSN-16ex in different bands are estimated and listed in Table~\ref{tab_lc_parameter}.  The absolute $V$ band light curves of these SNe  are displayed in Fig.~\ref{fig_Vabs_comp} and compared with other well-studied SNe Ia. 
SN 2013bz and ASASSN-16ex have luminosities comparable to SNe 2006gz/91T, ASASSN-15pz.  The light curve of SN 2013bz is similar to SN 1991T. With a constant difference of $\sim$ 0.4\,mag, the absolute $V$ band light curve of ASASSN-16ex runs almost parallel to SN 2009dc. The light curve of  PSN J0910+5003 is fainter than 09dc--like objects but brighter than normal SNe 2003du/05cf/11fe. 
At maximum PSN J0910+5003 is fainter than ASASSN-16ex. However, during the late phase, they show  similar light curve evolution.

\subsection{Bolometric Light curve}

The quasi-bolometric luminosities of SN 2013bz, PSN J0910+5003 and ASASSN-16ex were derived using the observed $UBVRI$ magnitudes in Table~\ref{tab_sn_mag}. The $UBVRI$ magnitudes were dereddened with the reddening values estimated in section \ref{sec_reddening} and converted to monochromatic fluxes using zero points from \citet*{bess98}. The derived fluxes on each night were integrated over the observed wavelength range to get the total flux at optical bands. The integrated flux is converted to quasi-bolometric luminosity using the distances derived in section \ref{sec_abs_mag}. 

The peak quasi-bolometric luminosity for SN 2013bz obtained by  integration of optical fluxes is $\log L_\text{bol}^\text{max}$ = 43.30\,erg\,s$^{-1}$. Adding 20\% flux to account for the contribution from the missing bands \citep{wang09} results in $\log L_\text{bol}^\text{max}$ = 43.38 $\pm$ 0.07\,erg\,s$^{-1}$. 

The maximum quasi-bolometric luminosity for ASASSN-16ex is obtained as $\log L_\text{bol}^\text{max}$ = 43.29\,erg\,s$^{-1}$ using optical observations. We used {\it Swift} UVOT data of ASASSN-16ex to estimate  contributions from UV bands.
The UV magnitudes  were dereddened following \citet{brow10} and converted to monochromatic flux using the zero points from \citet{pool08}. With the UV contribution, the peak bolometric luminosity of ASASSN-16ex becomes $\log L_\text{bol}^\text{max}$ = 43.39\,erg\,s$^{-1}$. The UV contribution to the bolometric luminosity is found as $\sim$ 29\%, 22\%, and 13\% at $-$7\,d, maximum, and +20\,d, respectively. The 09dc--like SNe Ia are found to be UV bright \citep{brow14,chak14,chen19,dimi22},  ASASSN-16ex follows a similar trend. After adding 5\% contribution from NIR bands \citep{wang09,scal10,yama16}, the peak bolometric luminosity of ASASSN-16ex is estimated as $\log L_\text{bol}^\text{max}$ = 43.40 $\pm$ 0.06\,erg\,s$^{-1}$.

The peak quasi-bolometric luminosity for PSN J0910+5003 is derived as $\log L_\text{bol}^\text{max}$ = 43.17\,erg\,s$^{-1}$
by integrating the optical data points. 
For 09dc--like objects, close to the maximum, the UV band contributes $\sim$ 20\% \citep{chak14} to the bolometric flux. We found similar UV contribution in ASASSN-16ex. On adding 20\% UV and 5\% NIR contributions, the peak bolometric luminosity of 
PSN J0910+5003 is obtained as $\log L_\text{bol}^\text{max}$ = 43.26 $\pm$ 0.07\,erg\,s$^{-1}$.

\begin{figure}
\includegraphics[width=\columnwidth]{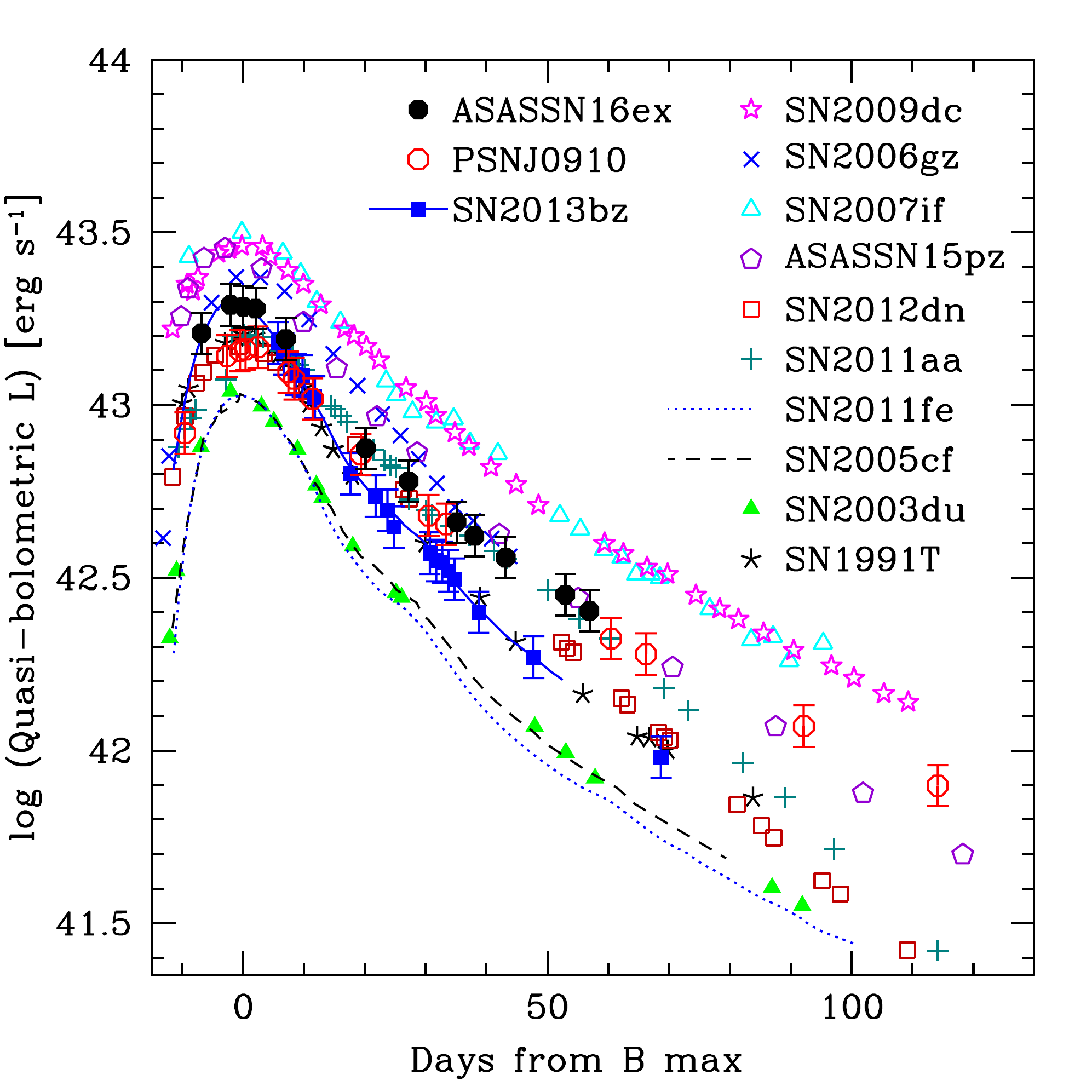}
\caption{Quasi-bolometric light curves of SN 2013bz, PSN J0910+5003 and ASASSN-16ex are plotted along with those of other well-studied SNe Ia.}
\label{fig_bol_curve}
\end{figure}

The quasi-bolometric light curves of SN 2013bz, PSN J0910+5003 and ASASSN-16ex are plotted in Fig.~\ref{fig_bol_curve} and compared with other well-studied SNe Ia. The peak quasi-bolometric luminosities of SN 2013bz, PSN J0910+5003 and ASASSN-16ex are lower than 09dc--like SNe 2006gz/07if/09dc/ASASSN-15pz and brighter than normal SNe Ia.   

\subsection{Mass of Nickel synthesized}
Type Ia SNe  are powered by the radioactive decay of $^{56}$Ni to $^{56}$Co and subsequently to $^{56}$Fe. The peak bolometric luminosity can be used to estimate the mass of $^{56}$Ni synthesized in the explosion of SNe Ia. 

The mass of $^{56}$Ni synthesized in the explosion of SN 2013bz, PSN J0910+5003 and ASASSN-16ex is estimated using Arnett's rule \citep{arne82}. This rule states that the peak bolometric luminosity of a type Ia SN is proportional to the instantaneous energy release rate from radioactive decay. This can be written as-
\begin{equation*}
\text{M}_\text{Ni} = \frac{L_\text{bol}^\text{max}}{\alpha \dot{S}(t_\text{R})}
\end{equation*}
where M$_\text{Ni}$ is the mass of $^{56}$Ni, $\alpha$ is the ratio of bolometric to radioactive luminosities (near unity), and $\dot{S}(t_\text{R})$ is the radioactivity luminosity per unit nickel mass  evaluated for the rise time $t_\text{R}$. From  \citet{nady94}, $\dot{S}(t_{R})$ can be written as-
\begin{eqnarray*}
\dot{S}(t_\text{R}) = \Bigl (6.45~e^{-(t_\text{R}/8.8\text{d})} + 1.45~e^{-(t_\text{R}/111.3\text{d})}\Bigr ) \\ 
\times   10^{43} ~~ \text{erg\,s}^{-1}\text{\,M}_\odot^{-1}
\end{eqnarray*}
where 8.8 and 111.3\,d are  {\it e}-folding lifetimes ($\tau$) of  $^{56}$Ni and $^{56}$Co, respectively.

With the lack of early observations, it is difficult to constrain these events' explosion epoch and rise time. This limits an accurate determination of the mass of $^{56}$Ni.
Rise time  ($t_\text{R}$) for the normal SNe Ia is found to be 17.4\,d \citep{hayd10}, 18\,d \citep*{gane11}, 
18.9\,d \citep{firt15,mill20}, 19.1 -- 19.6\,d \citep{ries99,conl06}. 
Using $t_\text{R}$ = 18 $\pm$ 2\,d, and $\alpha$ = 1.2 $\pm$ 0.2 \citep{bran92}, the mass of $^{56}$Ni synthesized in the explosion of SN 2013bz is estimated as M$_\text{Ni}$ = 0.96 $\pm$ 0.24\,M$_\odot$.
%(for SN 2013bz), M$_\text{Ni}$ = 0.74 $\pm$ 0.18\,M$_\odot$ (for PSN J0910+5003) and M$_\text{Ni}$ = 1.0 $\pm$ 0.24\,M$_\odot$ (for ASASSN-16ex).
The 09dc--like SNe are found to have longer $t_\text{R}$ $\sim$ 21 -- 24\,d \citep{scal10,silv11,chen19,luj21} with an average of $t_\text{R}$ $\sim$ 22 $\pm$ 4\,d \citep{asha21}. %The higher $t_\text{R}$ will give a larger value of M$_\text{Ni}$.
Using $t_\text{R}$ = 22\,d and $\alpha$ = 1.2, the mass of $^{56}$Ni in 
PSN J0910+5003 and ASASSN-16ex is estimated as 
M$_\text{Ni}$ = 0.89 $\pm$ 0.24\,M$_\odot$ and M$_\text{Ni}$ = 1.2 $\pm$ 0.32\,M$_\odot$, respectively.

\section{SPECTRAL EVOLUTION}
\label{spectral_evolution}
Spectral series were obtained for SN 2013bz, PSN J0910+5003 and ASASSN-16ex using HFOSC-HCT. In total, 11 spectra spanning +4 to +35\,d for SN 2013bz, 8 spectra from $-$3 to +63\,d for PSN J0910+5003, and 8 spectra from $-$8 to +43\,d for ASASSN-16ex were obtained{\footnote{Phase with respect to $B$ band maximum.}}. The details of spectroscopic observations are given in Table~\ref{tab_spec_log}. Spectral evolution of SN 2013bz, PSN J0910+5003 and  ASASSN-16ex are  presented in Fig.~\ref{fig13bz_spec}, Fig.~\ref{figpsn09_spec} and Fig.~\ref{figasn16ex_spec}, respectively. All the spectra are reddening and redshift corrected. Telluric lines are not removed. 

Spectral features in the early phase of  type Ia SNe are mostly absorption lines of singly ionized IMEs such as O, Mg, Si, S and Ca produced at outer photospheric layers. As the ejecta expands, the photosphere moves into deeper layers, and lines from the inner layers are seen. During this, features from IMEs are replaced by the IGEs, e.g., Fe, Co and Ni. 

\subsection{SN 2013bz}

\begin{figure}
\includegraphics[width=\columnwidth]{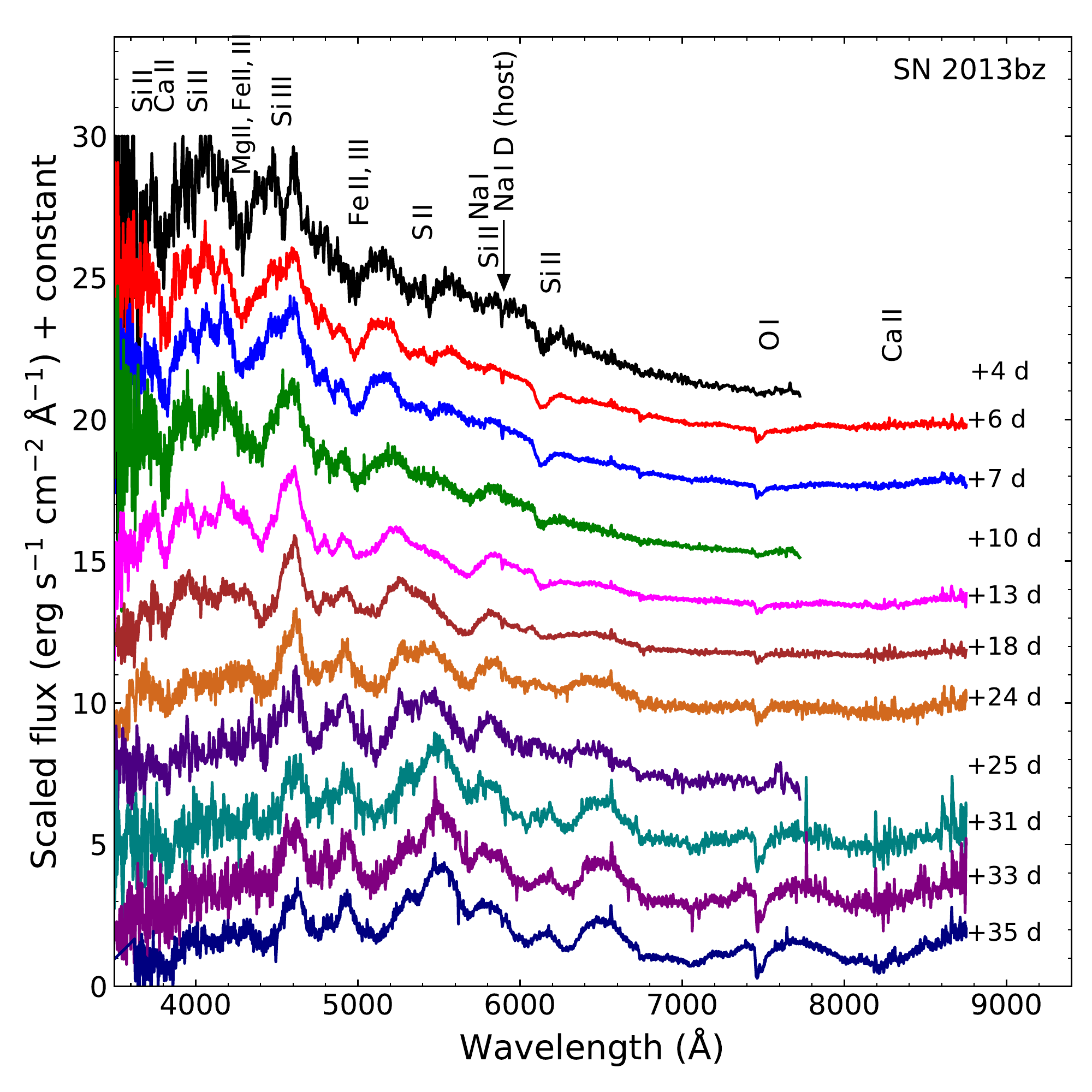}
\caption{Spectral evolution of SN 2013bz from $\sim$ +4 to +35\,d.}
\label{fig13bz_spec}
\end{figure}

The early post-maximum spectra of SN 2013bz (+4, +6 and +7\,d) show the characteristics features  of SNe Ia, mostly from IMEs marked in Fig.~\ref{fig13bz_spec}.  At the bluer end, a deep absorption seen around 4000\,\AA\ is due to Ca\,{\sc ii}\,H\&K lines. The next prominent and broad feature at around 4500\,\AA\ is due to a blend of Fe\,{\sc iii}\,$\lambda$4404, Mg\,{\sc ii}\,$\lambda$4481 and Fe\,{\sc ii}\,$\lambda$4555. Moving red-ward, a sharp absorption due to Si\,{\sc iii}\,$\lambda$4560 is seen in the first spectrum at +4\,d, which has become weak in the following spectrum at +6\,d. The Si\,{\sc iii}\,$\lambda$4560 is generally seen during the early hot photospheric phase. 
A broad blend is seen at $\sim$ 5000\,\AA\ due to Fe\,{\sc ii}\,$\lambda$$\lambda$4924, 5018, Si\,{\sc ii}\,$\lambda$5051, Fe\,{\sc iii}\,$\lambda$5129 and Fe\,{\sc ii}\,$\lambda$5169. The `W'-shaped S\,{\sc ii}\,$\lambda$$\lambda$5654, 5468 feature is also visible with its decreasing strength in subsequent spectra.  Features on the red side of the spectrum, such as O\,{\sc i} and Ca\,{\sc ii} NIR, are weak in SN 2013bz.
A small narrow feature seen at $\sim$ 6000\,\AA\ is due to Na\,{\sc i}\,D from the host galaxy. 

In the spectra obtained at +10, +13, and +18\,d, most of the 
 features due to IMEs'  such as `W'-shaped S\,{\sc ii} and Si\,{\sc ii}  weakened/disappeared. 
The features due to IMEs  are getting replaced by Fe\,{\sc ii} lines. Development of emission peaks,  such as at $\sim$ 4600\,\AA, \ can also be noticed. A strong absorption due to Na\,{\sc i} from  SN ejecta is visible at $\sim$ 5700\,\AA. The  spectra after one-month post-maximum are dominated  by Fe\,{\sc ii} lines. Feature due to Ca\,{\sc ii} NIR triplet looks stronger than earlier. 

\subsection{PSN J0910+5003}

\begin{figure}
\includegraphics[width=\columnwidth]{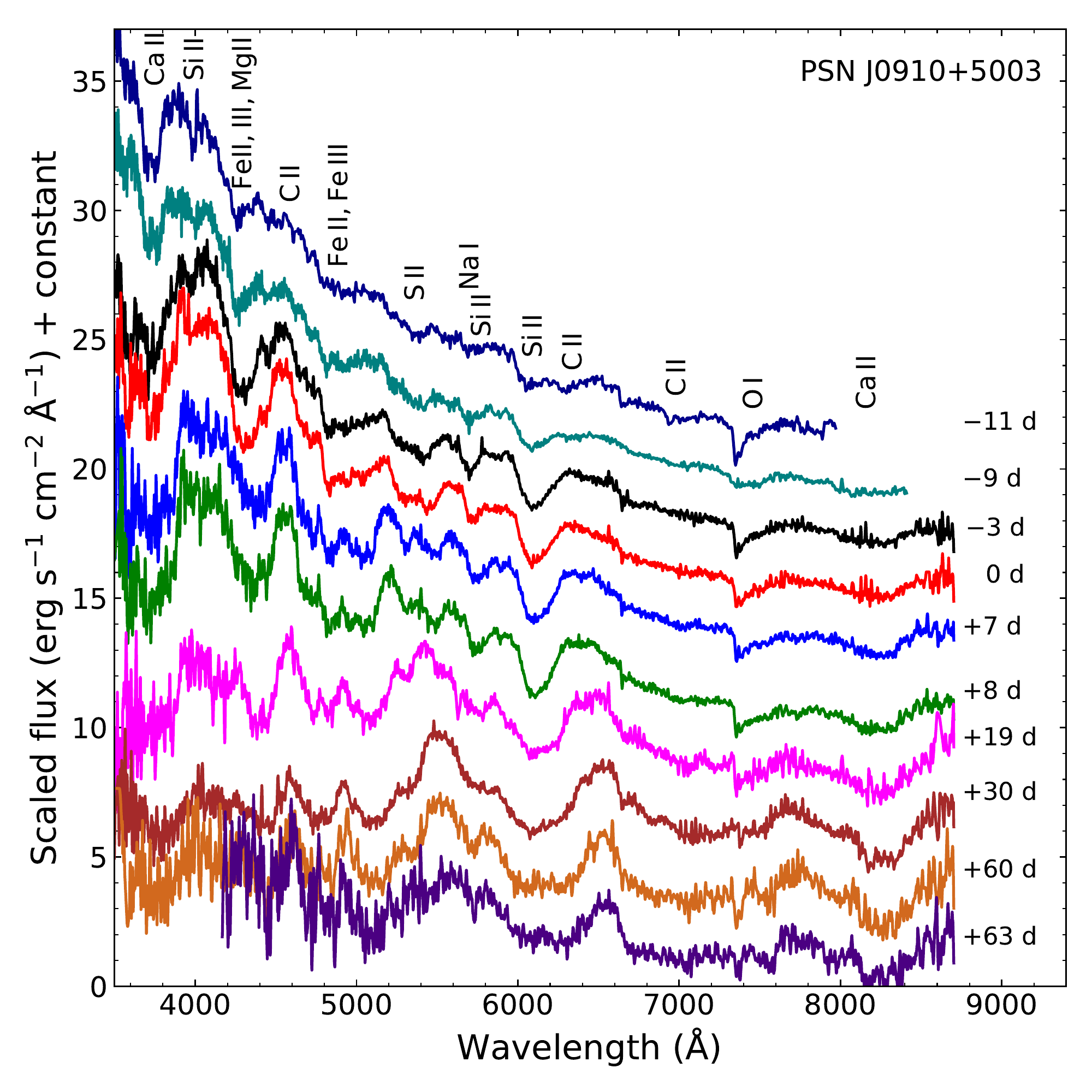}
\caption{Spectral evolution of PSN J0910+5003 from $\sim -$11 to +63\,d.}
\label{figpsn09_spec}
\end{figure}

The spectral evolution of PSN J0910+5003 from $-$11 to +63\,d is shown in Fig.~\ref{figpsn09_spec}. The first two spectra at $-$11 and $-$9\,d are taken from  WISeREP \citep{toma15,stah20} and displayed to show early evolution and spectral characteristics. The first spectrum at $-$11\,d looks like a featureless continuum; most spectral lines are shallow/not yet fully developed. 
  A strong C\,{\sc ii}\,$\lambda$6580 absorption can be seen in the red-wing of Si\,{\sc ii}\,$\lambda$6355, which makes the absorption at $\sim$ 6200\,\AA \ very broad. This indicates the presence of unburned material in the ejecta,  
usually found in 09dc--like SNe Ia \citep{taub11,chak14,chen19,luj21,jian21,dimi22,sriv23}. 

In the spectrum at $-$9\,d, the features from IMEs are strengthening. The carbon lines have almost disappeared; only a tiny suppression is seen at the red-wing of  Si\,{\sc ii}\,$\lambda$6355. By $-$3\,d, the absorption features due to IMEs are fully developed and dominate  the  spectra till one-week post-maximum. In contrast to SN 2013bz, features like  
Si\,{\sc ii}\,$\lambda$5972, O\,{\sc i} and Ca\,{\sc ii} NIR are stronger in PSN J0910+5003. 
Spectrum at +19\,d and subsequent spectra show the appearance of several emission peaks due to IGEs. Fe\,{\sc ii} lines have replaced the features from IMEs. 

\subsection{ASASSN-16ex}

\begin{figure}
\includegraphics[width=\columnwidth]{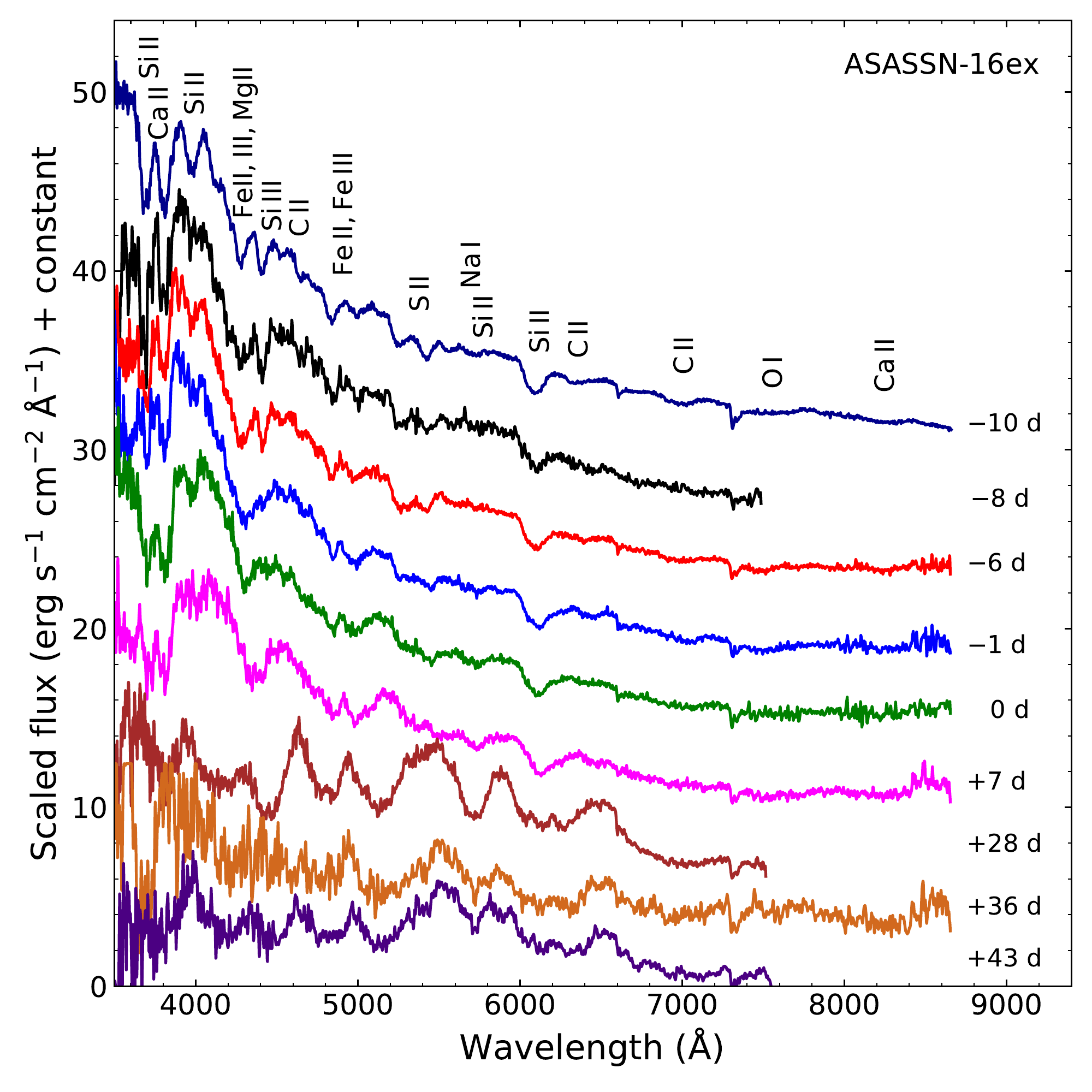}
\caption{Spectral evolution of ASASSN-16ex from $\sim -$10 to +43\,d.}
\label{figasn16ex_spec}
\end{figure}

The spectral evolution of ASASSN-16ex from $-$10 to +43\,d is shown in Fig.~\ref{figasn16ex_spec}. The first spectrum at $-$10\,d is taken from WISeREP \citep{toma16}. Similar to PSN J0910+5003, the signature of unburned C\,{\sc ii} is seen in this object. 
Compared to  PSN J0910+5003, spectral features in ASASSN-16ex are well developed and sharp. 
The  Si\,{\sc ii}\,$\lambda$5972, O\,{\sc i} and Ca\,{\sc ii} NIR features are weak in ASASSN-16ex than PSN J0910+5003.
Along with a weak Si\,{\sc ii}\,$\lambda$5972 line, a strong Si\,{\sc iii}\,$\lambda$4560 feature is present in the spectrum of ASASSN-16ex, suggesting  a hot photosphere. In the spectrum at $\sim$ one month and in the subsequent spectra, the features of IMEs are replaced by Fe\,{\sc ii} lines, and the development of emission lines can be seen.  

\subsection{Spectral Comparison}
\label{sec_spec_comp}

\begin{figure}
\includegraphics[width=\columnwidth]{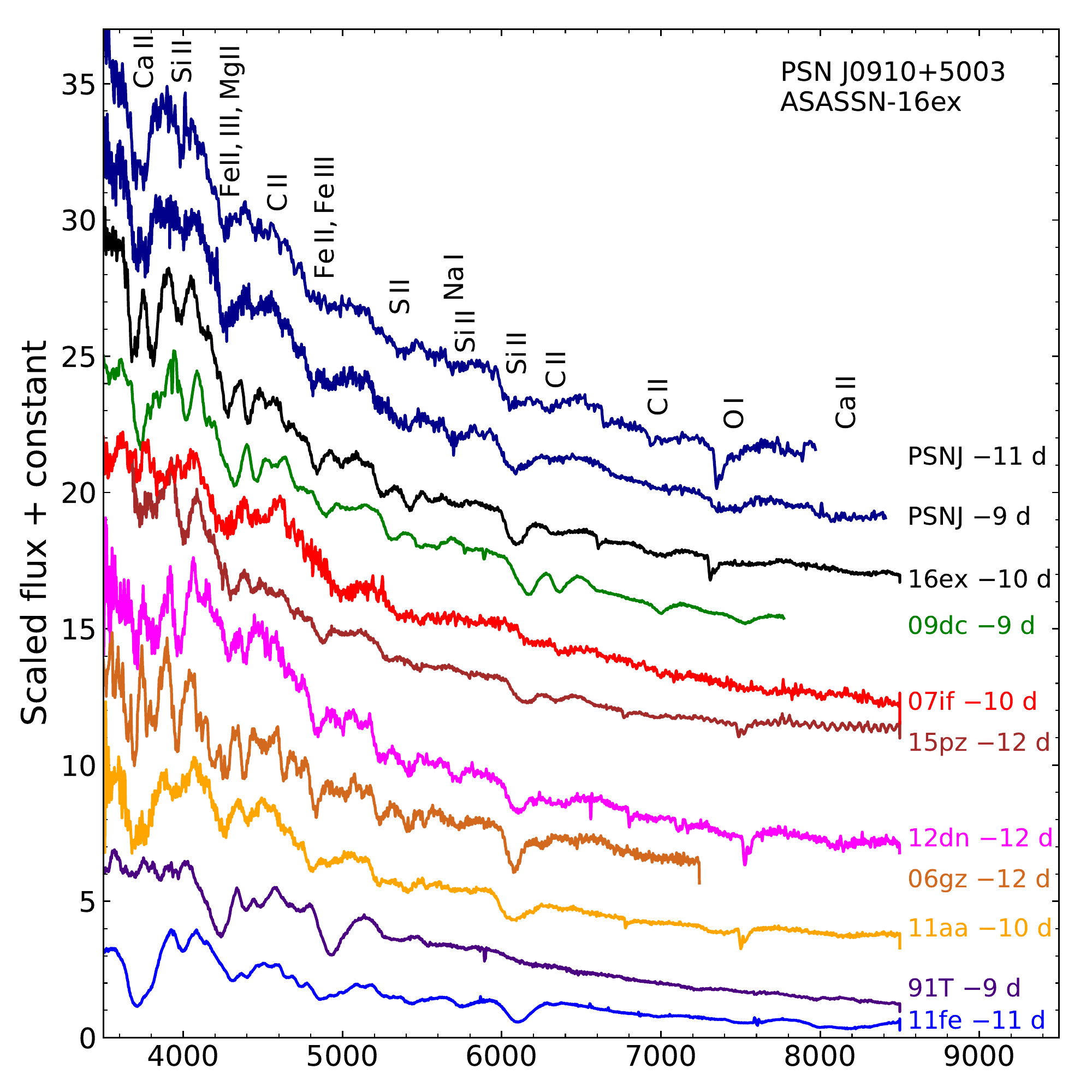}
\includegraphics[width=\columnwidth]{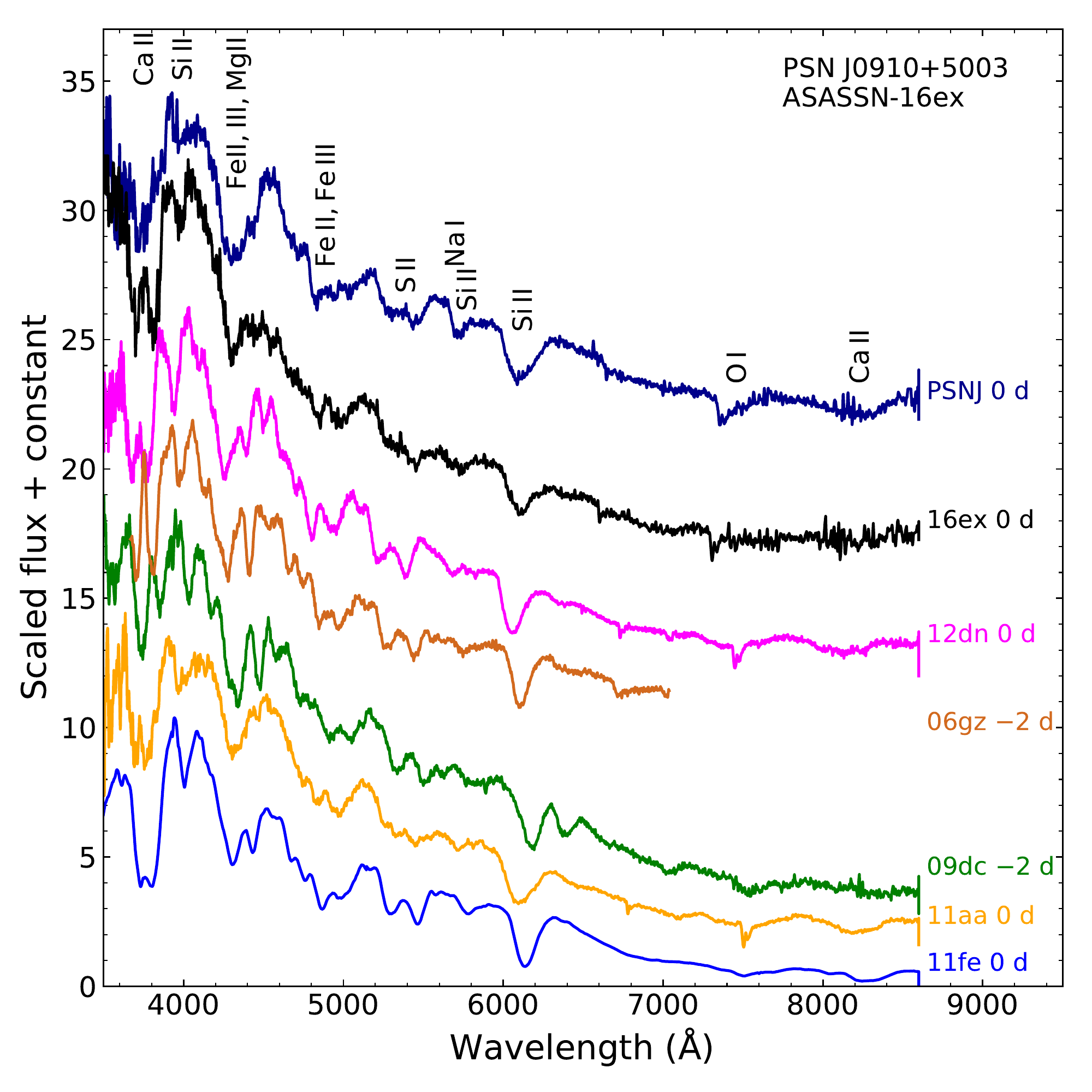}
\caption{Spectra of PSN J0910+5003 and ASASSN-16ex at $\sim -$ 10\,d ({\it top}) and at  maximum ({\it bottom}) are compared with other 
well-studied SNe Ia. }
\label{fig_spec_comp_m9}
\end{figure}

In Fig.~\ref{fig_spec_comp_m9} (top panel), the early spectra of PSN J0910+5003 ($-$11, $-$9\,d) and ASASSN-16ex ($-$10\,d) are compared with those of SN 1991T \citep{fili92}, SN 2006gz \citep{hick07}, SN 2007if \citep{scal10}, SN 2009dc \citep{taub11}, SN 2011fe \citep{parr12}, SN 2011aa \citep{dutt22} and ASASSN-15pz \citep{chen19} at similar epoch.
Though the lines in PSN J0910+5003 are shallower, the spectra of both PSN J0910+5003 and ASASSN-16ex are similar to 09dc--like SNe.
Strong C\,{\sc ii} features are seen in both events. 

The spectral features of ASASSN-16ex are very similar to those of 2006gz/09dc/12dn.
In the  bluer region, well-separated sharp absorption due to  Si\,{\sc ii}\,$\lambda$3858, Ca\,{\sc ii}\,H\&K, Si\,{\sc ii}\,$\lambda$4130 and Si\,{\sc iii}\,$\lambda$4560  are identical in these events.  
Other features like  `W'-shaped S\,{\sc ii}, Si\,{\sc ii}\,$\lambda$6355, and a shallower Si\,{\sc ii}\,$\lambda$5972 look similar in these objects. The spectral features in PSN J0910+5003 at $-$11 and $-$9\,d are similar to those in SN 2007if and SN 2011aa. 
A comparison of spectra of PSN J0910+5003, ASASSN-16ex and other SNe is  made at the maximum phase in the bottom panel of Fig.~\ref{fig_spec_comp_m9}. At this phase also, ASASSN-16ex closely resembles SN 2012dn/06gz. The spectrum of PSN J0910+5003 is similar to SN 2011aa.
Absorption lines in PSN J0910+5003 appear broader and blended.

\begin{figure}
\includegraphics[width=\columnwidth]{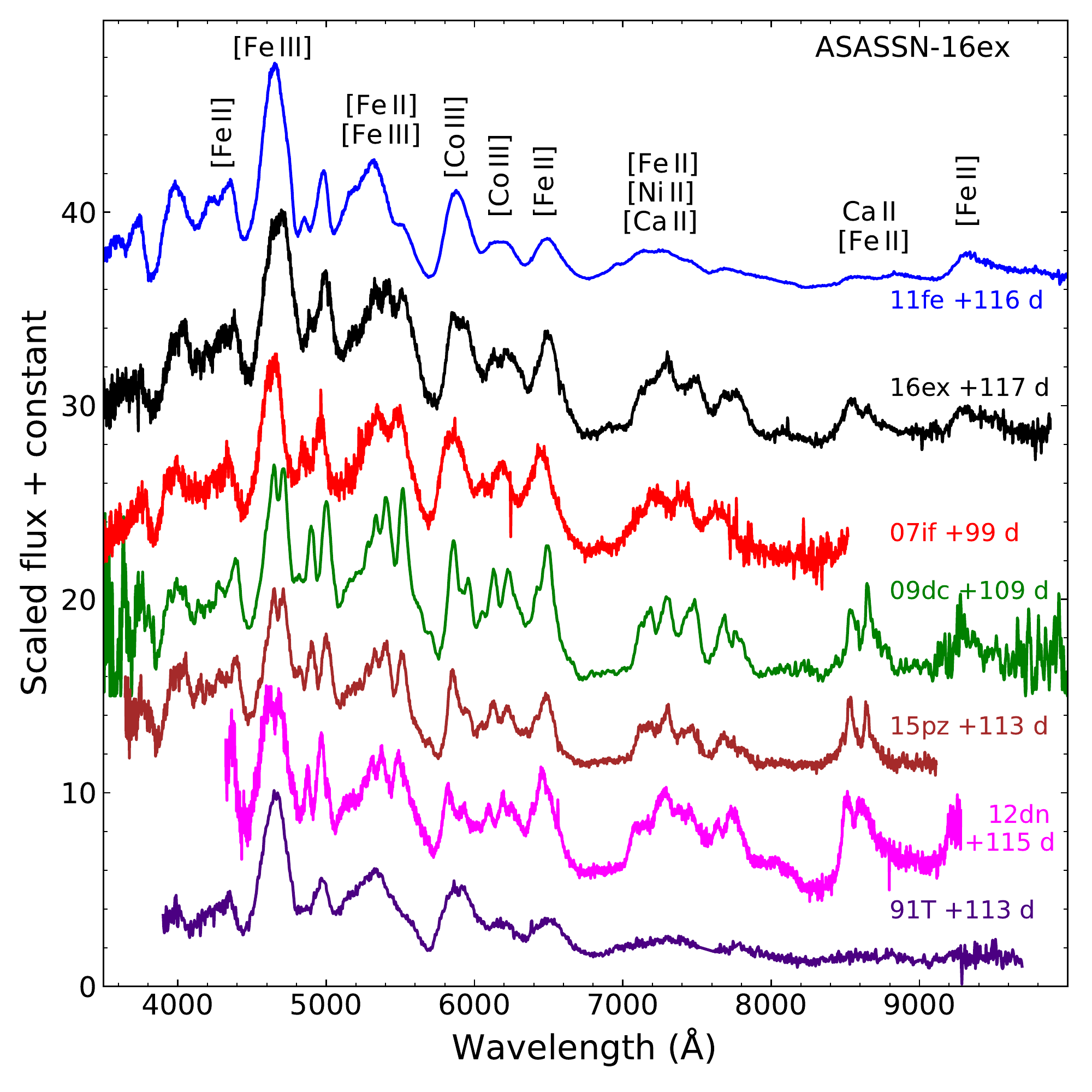}
\caption{The spectrum of ASASSN-16ex at +117\,d is compared with other well-studied SNe Ia. }
\label{fig_spec_comp_p117}
\end{figure}

In Fig.~\ref{fig_spec_comp_p117}, the spectrum of ASASSN-16ex  $\sim$ four months after maximum light (taken from WISeREP, \citealt{stah20}) is compared with those of SN 2011fe \citep{stah20}, SN 2007if \citep{silv11}, SN 2009dc \citep{taub11}, ASASSN-15pz \citep{chen19}, SN 1991T \citep{silv12_spec1} and SN 1991bg \citep{tura96} at a similar epoch.  
A few months after the explosion, SN enters into nebular phase, the ejecta becomes optically thin, and the inner part of the ejecta becomes  visible. Forbidden emission lines of IGEs, i.e., Fe, Co, and Ni, characterize spectra of SNe Ia at this phase. Some prominent nebular features are marked in Fig.~\ref{fig_spec_comp_p117}.  Similarity in the spectral features between ASASSN-16ex and  a normal SN Ia can be seen. However, there are certain noticeable differences also. The blending of lines in ASASSN-16ex is less than a normal SN Ia. In the red region beyond $\sim$ 6500\,\AA, features in ASASSN-16ex are more pronounced as compared to normal SN 2011fe and luminous SN 1991T. There are a number of similarities between the spectrum of ASASSN-16ex and 09dc--like SNe, which include 
(i) double-peaked [Fe\,{\sc ii}]+[Fe\,{\sc iii}] complex at $\sim$ 5400\,\AA; strength of [Fe\,{\sc iii}] 4700\,\AA\ appears lower relative to [Fe\,{\sc ii}]+[Fe\,{\sc iii}] 
(ii) strong [Fe\,{\sc ii}] at $\sim$ 6500\,\AA\ 
(iii) a broad absorption trough with a flat bottom around 6700 -- 7000\,\AA\ (iv) multiple emission peaks around 7000 -- 7900\,\AA\ 
(v) sharp emission peaks at $\sim$ 8500\,\AA. 

\subsection{SYN++ Synthetic model spectra}
\label{sec_synow}

\begin{figure}
\includegraphics[width=\columnwidth]{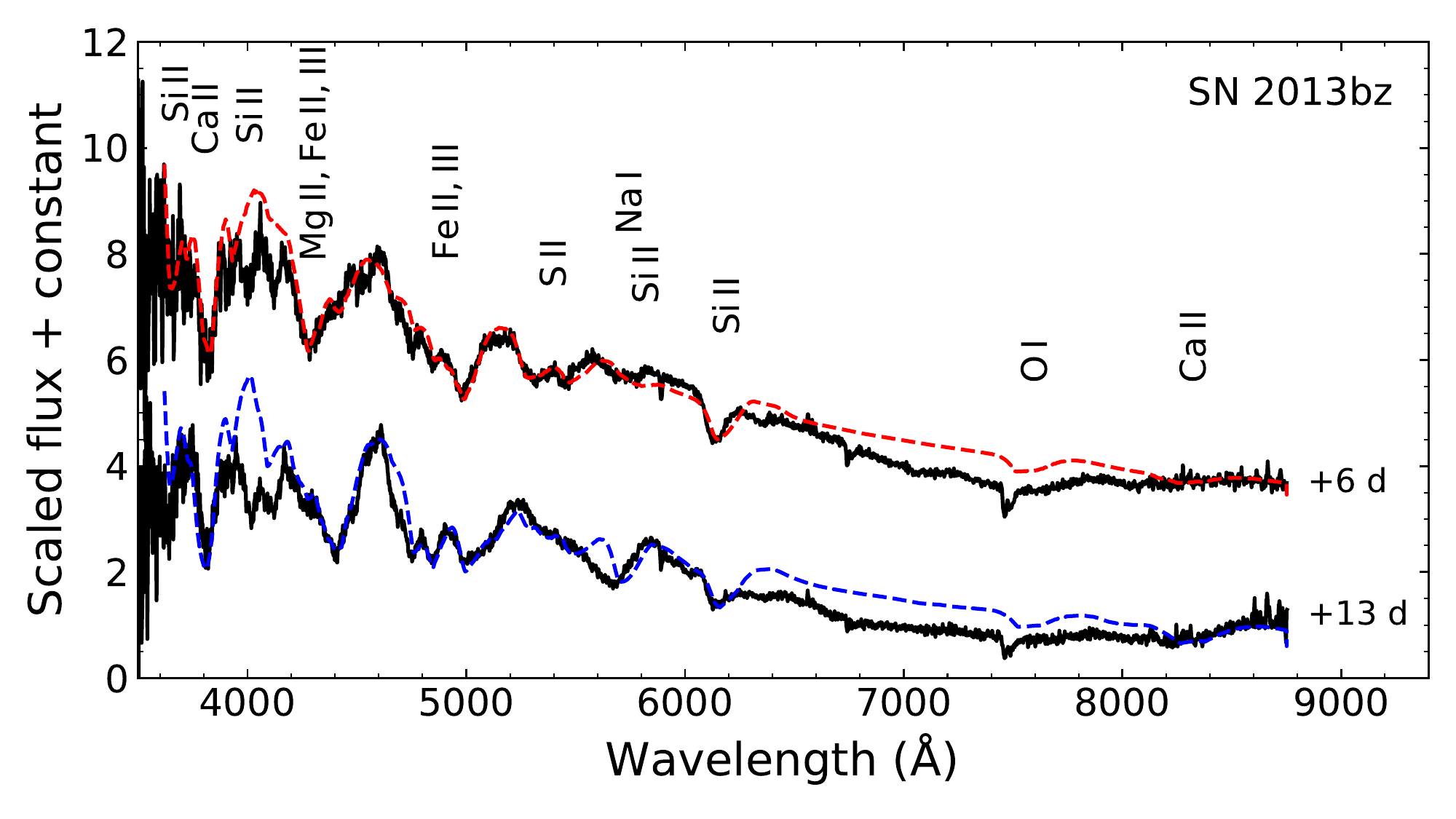}
\caption{The synthetic spectra generated using
SYN++ code are compared with the observed spectra of SN 2013bz.}
\label{fig13bz_synow}
\end{figure}

Spectra of SN 2013bz at +6 and +13\,d are compared  with the
synthetic spectra generated using  SYN++ code \citep*{fish00,thom11} and plotted in Fig.~\ref{fig13bz_synow}. The observed spectrum at +6\,d matches the synthetic spectrum with  a photospheric velocity ($v_{ph}$) of 10\,500\,km\,s$^{-1}$ and blackbody temperature ($T_{BB}$) of 11\,000\,K. The species used are marked. 
The spectrum at +13\,d matches with similar $T_{BB}$ and species but with a lower $v_{ph}$ of 10\,200\,km\,s$^{-1}$.

\begin{figure}
\includegraphics[width=\columnwidth]{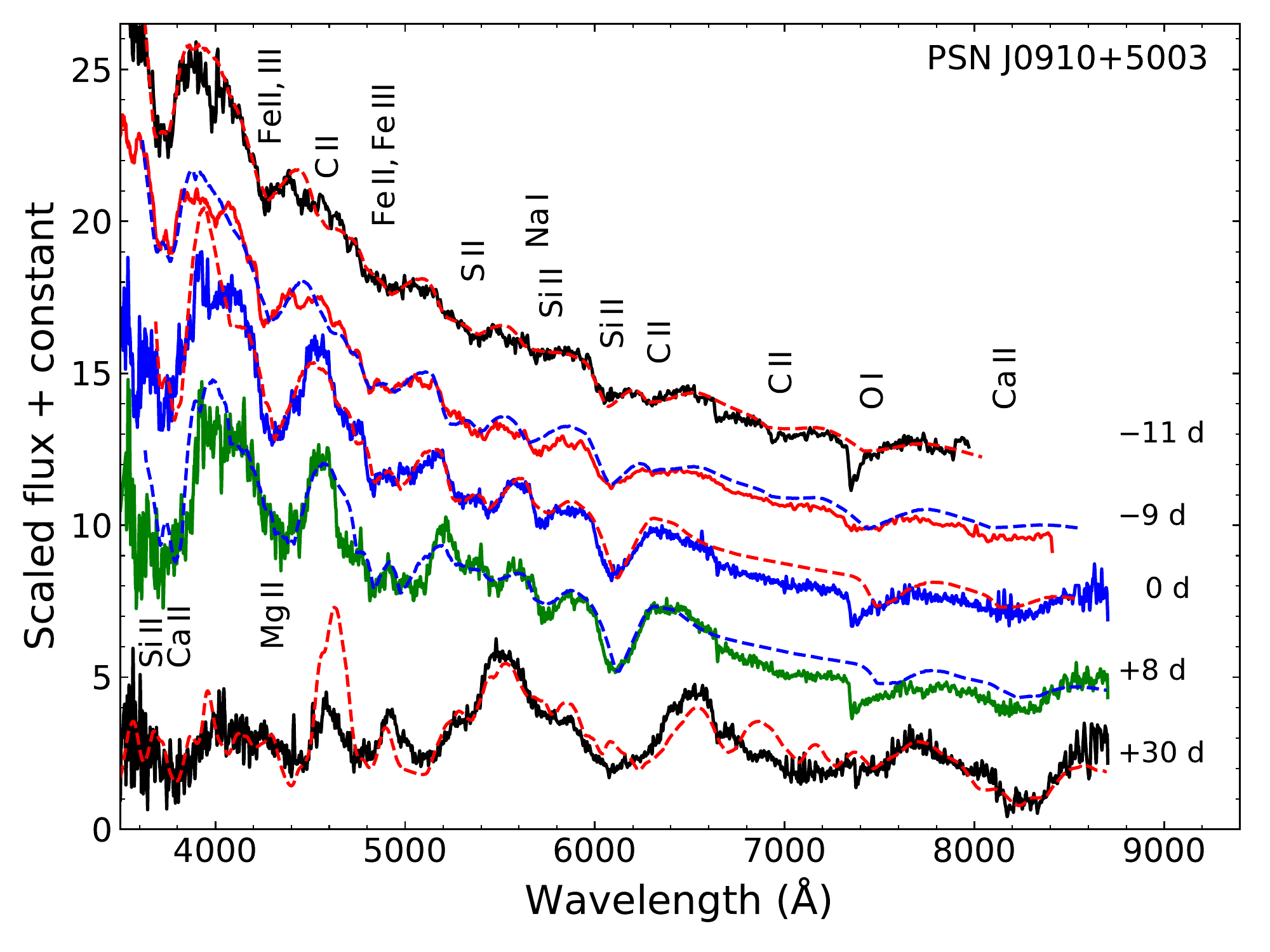}
\caption{The synthetic spectra generated using SYN++ code are compared with the observed spectra of PSN J0910+5003. }
\label{figpsn09_synow}
\end{figure}

Spectra of PSN J0910+5003 at $-$11, $-$9, 0, +8 and +30\,d are matched with the synthetic spectrum and plotted in Fig.~\ref{figpsn09_synow}. The synthetic spectrum at $-$11\,d is produced using $v_{ph}$ = 14\,500\,km\,s$^{-1}$ and $T_{BB}$ =  15\,000\,K. Ions used are marked in the figure.
The synthetic spectrum at $-$9\,d is reproduced with similar species, $v_{ph}$ = 13\,500\,km\,s$^{-1}$ and $T_{BB}$ = 14\,300\,K. 
For the maximum and one-week post-maximum phase, we used $v_{ph}$ =  11\,700 and 11\,500\,km\,s$^{-1}$, $T_{BB}$ = 13\,000 and 11\,500\,K, respectively. Species used are similar to the $-$9\,d spectrum;  C\,{\sc ii} is not included. The synthetic spectrum at +30\,d has a $v_{ph}$ =  10\,000\,km\,s$^{-1}$ and $T_{BB}$ = 8\,000\,K. It includes species of Si\,{\sc ii}, Ca\,{\sc ii}, Fe\,{\sc ii}, Ni\,{\sc ii} and Co\,{\sc ii}. 

\begin{figure}
\includegraphics[width=\columnwidth]{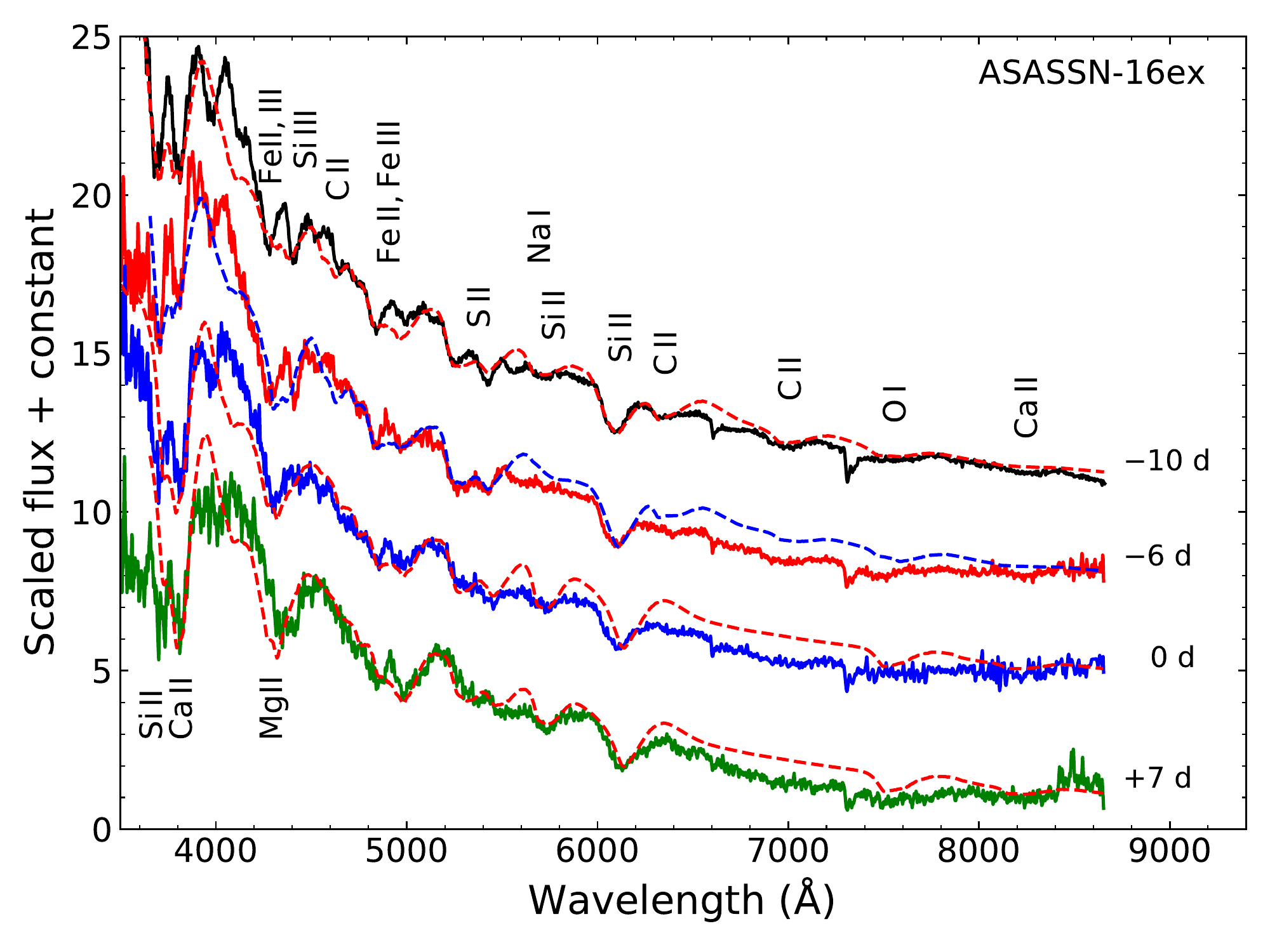}
\caption{The synthetic spectra generated using SYN++ code are compared with the observed spectra of ASASSN-16ex. }
\label{figasn16ex_synow}
\end{figure}

Spectra of ASASSN-16ex at $-$10, $-$6, 0, and +7\,d are matched with the synthetic spectrum and plotted in Fig.~\ref{figasn16ex_synow}. The synthetic spectrum at $-$10\,d is generated using   $v_{ph}$ = 12\,500\,km\,s$^{-1}$ and $T_{BB}$ =  15\,000\,K. Species used are marked. 
The observed spectrum of ASASSN-16ex at $-$6\,d matches the synthetic spectrum with  similar parameters and species used at $-$10\,d. The synthetic spectra at maximum and one-week post-maximum phase have lower $v_{ph}$ (11\,200\,km\,s$^{-1}$ and 11\,000\,km\,s$^{-1}$) and  $T_{BB}$ (14\,500 and 14\,000\,K).
Both include species of O\,{\sc i}, Na\,{\sc i}, Mg\,{\sc ii}, Si\,{\sc ii}, S\,{\sc ii}, Ca\,{\sc ii}, Fe\,{\sc ii}, and Fe\,{\sc iii}.

\subsection{Velocity and spectral parameters}
\label{sec_velocity}
\begin{figure}
\includegraphics[width=\columnwidth]{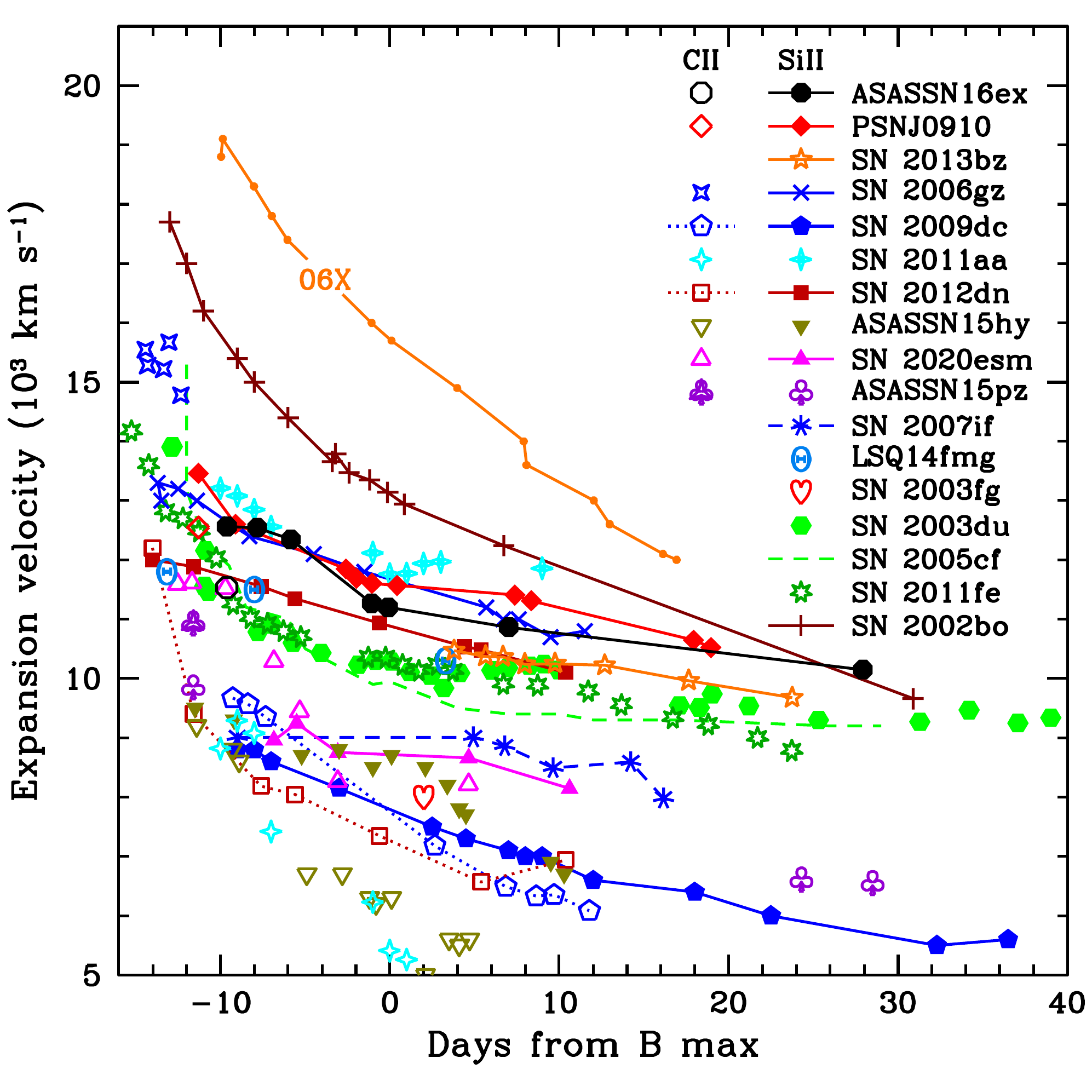}
\caption{Photospheric velocity evolution measured using Si\,{\sc ii}\,$\lambda$6355 absorption line for SN 2013bz, PSN J0910+5003 and ASASSN-16ex is compared with other well-studied SNe Ia. The velocity of the C\,{\sc ii}\,$\lambda$6580 line is also displayed for PSN J0910+5003 and ASASSN-16ex, along with other SNe Ia.} 
\label{fig_vel}
\end{figure}

The photospheric velocities of SN 2013bz, PSN J0910+5003 and ASASSN-16ex estimated using the absorption minimum of Si\,{\sc ii}\,$\lambda$6355  are plotted in Fig.~\ref{fig_vel}, along with other well-studied SNe Ia for comparison.
The velocity evolution of SN 2013bz is similar to normal event SN 2003du. The velocity evolution of  PSN J0910+5003 and ASASSN-16ex is  similar to SN 2006gz and the early evolution of SN 2011aa.
The Carbon $\lambda$6580 line,  detected in the early phase ($\sim -$10\,d) is found to have 
$\sim$ 1\,000\,km\,s$^{-1}$ lower velocity than that of the Si\,{\sc ii} line in both PSN J0910+5003 and ASASSN-16ex. This could be due to the clumping/line of sight effect \citep{parr11}. 

Normal SNe Ia are characterized by a pair of Si\,{\sc ii}\,$\lambda$5972 and Si\,{\sc ii}\,$\lambda$6355 features near the maximum phase. SNe Ia can be categorized and studied based on the various spectroscopic parameters derived from these lines, such as EWs \citep{bran06} and strength ratio $\cal R_\text{Si}$ \citep{nuge95}, the photospheric velocity measured from Si\,{\sc ii}\,$\lambda$6355 line \citep{wang09hv} and its gradient \citep{bene05}. We measured these parameters for SN 2013bz, PSN J0910+5003 and ASASSN-16ex, and listed in Table~\ref{tab_sp_param}. 
All three SNe belong to  normal velocity (NV, \citealt{wang09hv}) and low velocity gradient (LVG, \citealt{bene05}) subgroup. SN 2013bz and ASASSN-16ex fall in the shallow silicon (SS) subgroup, while PSN J0910+5003 is near the boundary of SS and core normal (CN) subgroup of SNe Ia \citep{bran06}. 

\begin{table}
\caption{Spectroscopic parameters of SN 2013bz, PSN J0910+5003 and ASASSN-16ex.}
\small
\centering
\setlength\tabcolsep{2pt} % default value: 6pt
\begin{tabular}{@{}lccc@{}}
\hline
Parameter$^a$ & 13bz & PSN & 16ex \\
\hline
1. $v_{ph}$ (km\,s$^{-1}$) & 10500 $\pm$ 600& 11600 $\pm$ 200& 11200 $\pm$ 400\\
2. $\dot{v}_\text{Si}$ (km\,s$^{-1}$\,d$^{-1}$) & 40 $\pm$ 3 & 31 $\pm$ 3  & 50 $\pm$ 4 \\
3. EW (Si\,{\sc ii}\,$\lambda$5972\,\AA)  & 11 $\pm$ 2& 20 $\pm$ 2 & 16 $\pm$ 2  \\
4. EW (Si\,{\sc ii}\,$\lambda$6355\,\AA)  & 30 $\pm$ 3 & 83  $\pm$ 8 & 60 $\pm$ 6 \\
5. $\cal R_\text{Si}$  & 0.34  $\pm$ 0.03 & 0.37  $\pm$ 0.04 & 0.32  $\pm$ 0.03 \\
6. Spectroscopic class$^b$ & LVG & LVG  & LVG  \\
7. Spectroscopic class$^c$ & SS & SS/CN  & SS  \\
8. Spectroscopic class$^d$ & NV & NV  & NV  \\
\hline
\multicolumn{4}{@{}l}{$^a$near $B$ max, $^b$\citet{bene05}} \\
\multicolumn{4}{@{}l}{$^c$\citet{bran06}, $^d$\citet{wang09hv}}
\label{tab_sp_param}
\end{tabular}
\end{table}

\section{Discussion and Summary}
\label{sec_summary}

SN 2013bz, PSN J0910+5003 and ASASSN-16ex were classified as 09dc--like objects using their early spectra
\citep{ochn13,toma15,toma16,pias16}. We have conducted a detailed photometric  and spectroscopic study on these objects to further investigate their properties. 
Among these, SN 2013bz is a slow-declining, luminous event with a decline rate parameter of $\Delta m_{15}(B)_{true}$ = 0.92 $\pm$ 0.04. Its photometric and spectral characteristics are similar to normal SNe Ia, while PSN J0910+5003 and ASASSN-16ex are similar to 09dc--like SNe Ia. The light curves of both PSN J0910+5003 and ASASSN-16ex are very broad relative to normal events. Their decline rate parameters ($\Delta m_{15}(B)_{true}$ = 0.70 $\pm$ 0.05 and 0.73 $\pm$ 0.03) are similar to 09dc--like SNe Ia. Further, ASASSN-16ex is very blue and bright in UV bands. 
The absolute luminosities of SN 2013bz, PSN J0910+5003 and ASASSN-16ex are on the higher side of normal SNe Ia. Their $B$ band peak absolute magnitudes are estimated as $-$19.61 $\pm$ 0.20\,mag, $-$19.44 $\pm$ 0.20\,mag and $-$19.78 $\pm$ 0.20\,mag, respectively. 

 The peak bolometric luminosities for these objects are derived as $L_{\text {bol}}^{\text {max}}$ = 43.38 $\pm$ 0.07\,erg\,s$^{-1}$, 43.26 $\pm$ 0.07\,erg\,s$^{-1}$ and 43.40 $\pm$ 0.06\,erg\,s$^{-1}$, respectively. The mass of $^{56}$Ni synthesized in the explosion of these events are estimated as 0.96 $\pm$ 0.24\,M$_\odot$, 0.89 $\pm$ 0.24\,M$_\odot$, and 1.2 $\pm$ 0.32\,M$_\odot$, respectively. The contribution of the UV flux in the bolometric luminosity for ASASSN-16ex is estimated to be 22\% at  $B$ maximum. The late phase spectrum of ASASSN-16ex, $\sim$ four months after maximum light, also looks very similar to those of 09dc--like SNe Ia.

\begin{figure}
\includegraphics[width=\columnwidth]{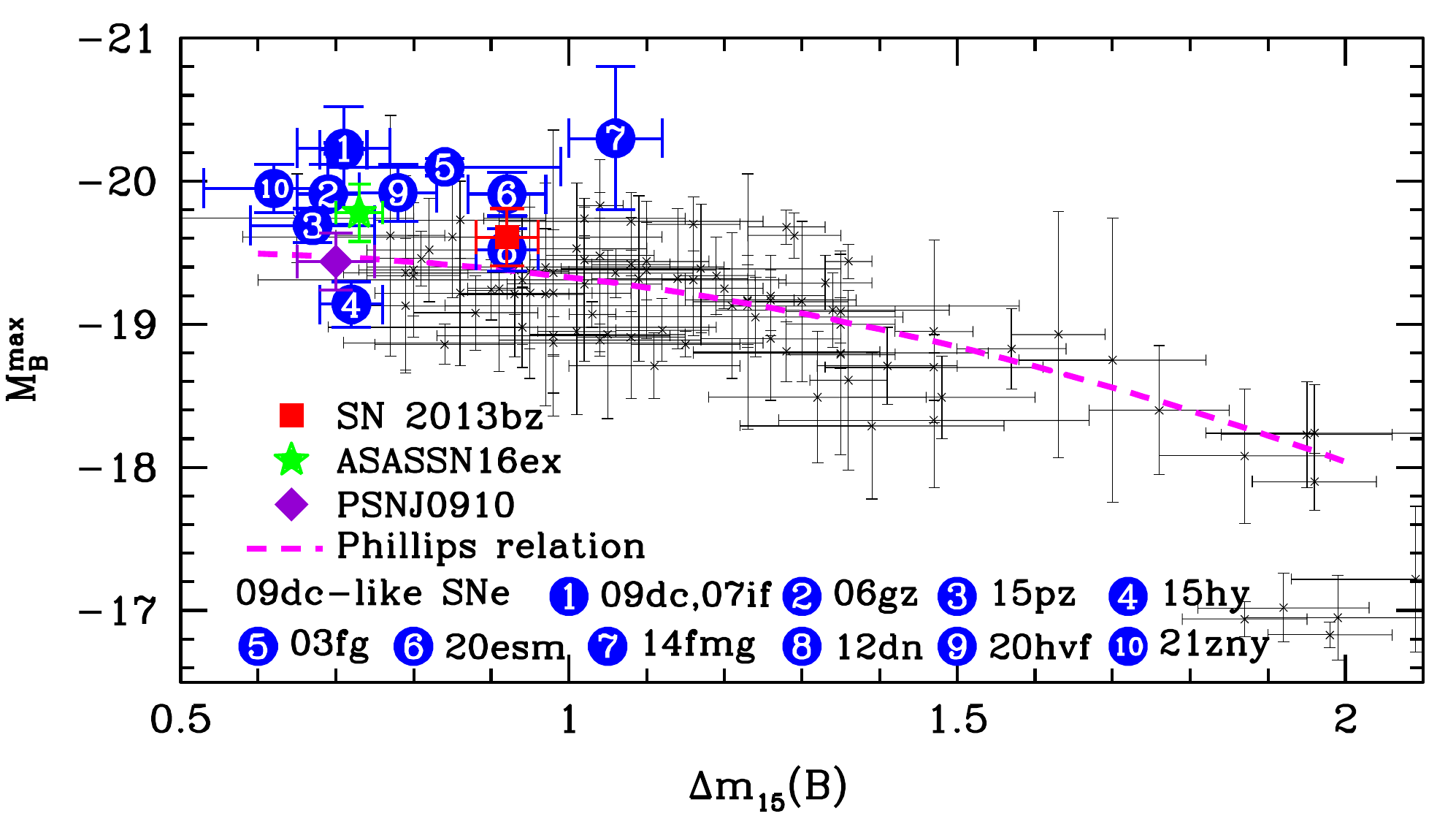}
\caption{Position of SN 2013bz, PSN J0910+5003 and ASASSN-16ex are shown in the {\it luminosity width relation} diagram of \citet{phil99} along with the known 09dc--like objects and a sample of type Ia SNe from \citet{hick09}. }
\label{fig_phillips_relation}
\end{figure}

Position of SN 2013bz, PSN J0910+5003 and ASASSN-16ex along with known 09dc--like objects and a sample of SNe Ia from \cite{hick09}, in the {\it luminosity width relation} diagram of \cite{phil99}, is shown in Fig.~\ref{fig_phillips_relation}.
\citet{phil99} used SNe Ia in the range of 0.85 $< \Delta m_{15}(B)_{true} < $ 1.7. Since then, the sample of SNe Ia has increased significantly with more number of objects in both the higher and lower luminosity end. Normal SNe Ia are suggested to follow this relation.

The 09dc--like SNe Ia are now known to have a large dispersion in their luminosity.
SNe 2003fg, 07if, 09dc, LSQ14fmg were exceptionally bright \citep{howe06,scal10,yuan10,yama09,silv11,taub11,hsia20}, while ASASSN-15hy had a lower luminosity \citep{luj21}.  Other SNe 2006gz, 12dn, ASASSN-15pz, 20esm, 20hvf and 21zny have luminosities in between \citep{hick09,maed09,chak14,taub19,chen19,dimi22,jian21,dimi23}. Their decline rate ranges from 0.62 $\pm$ 0.09 (SN 2021zny, \citealt{dimi23}) to 1.06 $\pm$ 0.06 (LSQ14fmg, \citealt{hsia20}). Except for LSQ14fmg, all the 09dc--like objects have 
$\Delta m_{15}(B)$ lower than a typical SN Ia. From Fig.~\ref{fig_phillips_relation}, it is clear that with a slower decline, PSN J0910+5003 and ASASSN-16ex lie towards the region occupied by 09dc--like objects and are moderately luminous.
SN 2013bz falls in the region of normal-luminous SNe Ia.

One of the characteristics seen in the 09dc--like SNe is the appearance of delayed $I$ band maximum. 
Timing of $I$ band peak relative to the $B$ band peak for 09dc--like SNe, 91T and normal SNe are listed in Table~\ref{tab_peak}. PSN J0910+5003 has a delayed $I$ band peak (6\,d), similar to other objects showing the most delayed $I$ band peak, e.g., SN 2006gz (6\,d), SN 2011aa (7\,d) and ASASSN-15hy (7\,d). However, ASASSN-16ex does not show the delayed $I$ band peak. 

\citet{kase06} suggested that double-peak appearance in the NIR light curves can be regarded as a direct consequence of the ejecta stratification and concentration of IGEs in the central regions. The secondary maximum results due to the ionization evolution of IGEs in the ejecta and transition from doubly ionized to singly ionized (Fe\,{\sc iii} to Fe\,{\sc ii}) state \citep{pint00,kase06}. Brighter SNe are expected to have a more pronounced and delayed secondary maximum but may vary depending on the amount and mixing of $^{56}$Ni, stable Fe/metallicity, and the abundance of Calcium \citep{kase06}. \citet{fola10} suggested that mixing  has the most significant effect on the strength of secondary maximum, and it can transform double-peak morphology into a single peak, as seen in the 09dc--like SNe.

\citet{asha21} suggested that full mixing in the ejecta should cause a prominent $H$ band break in the NIR spectrum, which is not seen during the early post-maximum phase  in the 09dc--like SNe;  instead, it is delayed by $\sim$ 1-2 months. The delayed appearance of the $H$ band break indicates that the $^{56}$Ni resides in the very inner layers. Vigorous mixing would bring $^{56}$Ni in the outermost region of the ejecta. However, the decay of surface $^{56}$Ni was not a good fit for the short duration early excess of SN 2020hvf \citep{jian21}. 

Explaining the bolometric light curve of 09dc--like SNe requires a higher $^{56}$Ni mass, suggesting ejecta mass in excess of Chandrasekhar mass \citep{scal10}. In the SD scenario, differentially rotating WDs can have mass over Chandrasekhar mass \citep{yoon05}, a possible progenitor for 09dc--like SN Ia \citep{howe06}. However, \citet{pfan10a}; \citet*{pfan10b}; \citet{hach12,fink18} show that the explosion energetics/burning products do not match well with this model. The deflagration model of differentially rotating WDs produces SNe Iax--like events. The explosion in detonation/delayed detonation is very powerful, the velocities of IMEs are significantly high, and most of the materials are burned to IGEs with just a small portion remaining unburned, contrary to what is seen in 09dc--like SNe Ia. 

The ejecta-CSM interaction has also been proposed as an alternative energy source to power these events. In this scenario,  09dc--like SNe are suggested to be a thermonuclear explosion in an enshrouded shell/envelope \citep*{khok93,hoef96} of C/O-rich materials with which the supernova ejecta interacts. The conversion/reprocessing  of kinetic energy/shock interaction energy acts as an additional source \citep{hick07,scal10,taub13,noeb16,hsia20,asha21,luj21}.

The signature of ejecta-CSM has been observed in some 09dc--like SNe Ia. In SN 2012dn, the excess NIR luminosity was interpreted due to the  presence of CSM dust (\citealt{yama16}; \citealt*{naga17}). Short-duration pulse-like early excess emission in SN 2020hvf \citep{jian21}, SN 2021zny \citep{dimi23} and SN 2022ilv \citep{sriv23}, within a few hours of the supernova explosion was modelled as an interaction of SN ejecta with the  CSM close to the progenitor \citep[see also][]{maed23}. However, the absence of  narrow emission lines in  the spectra of 09dc--like SNe Ia, indicates the CSM to be H/He-poor.

The origin of CSM is not well understood. \citet{yama16,naga17} suggest an SD scenario. The merger of two WDs in the DD scenario gives a high ejecta mass suitable for the 09dc--like SNe Ia \citep{hick07}, and naturally explains the absence of H/He features. During the disruption, a WD merging with another WD can form a dense envelope of C/O-rich materials. However, the physics of merging WDs needs to be understood, whether it is able to explain SNe Ia \citep{hoef96,frye10,shen12,moll14,rask14} or it will result in an accretion-induced collapse \citep{saio85}.
\citet{hsia20,luj21,asha21} suggest a `core-degenerate' scenario \citep{kash11}, where the explosion takes place in a degenerate C/O core enveloped by an AGB star. However, interaction signatures, e.g., bright X-ray emission, UV late-time rebrightening, narrow H/He lines, etc., have not been observed. 

SN 2011aa was found to be the slowest declining type Ia event with photometric evolution similar to 09dc--like objects. However, it was neither  over-luminous nor showed narrow spectral features, as seen in   09dc--like objects \citep{dutt22}. Photometric properties of PSN J0910+5003 closely match with the 09dc--like objects, while  spectral lines in PSN J0910+5003 are broad, similar to SN 2011aa (refer to Section~\ref{sec_spec_comp}). Early velocity evolution of PSN J0910+5003 is also similar to SN 2011aa (refer to Fig.~\ref{fig_vel}). \citet{dutt22} have shown that events like SN 2011aa could be explained by the violent merger of white dwarfs.

The strong C\,{\sc ii} feature is seen in the early spectra of PSN J0910+5003 and ASASSN-16ex at $\sim -$10\,d. The presence of strong carbon lines is a characteristic of 09dc--like objects. Though carbon is seen in early spectra of a good fraction of normal SNe Ia, it generally disappears soon \citep{parr11,fola12_carbon,silv12_carbon}. Some of the 09dc--like objects have shown carbon features extending up to maximum/post-maximum phase, e.g., SN 2009dc \citep{taub11}, SN 2012dn \citep{chak14,parr16,taub19}, LSQ14fmg \citep{hsia20}, ASASSN-15hy \citep{luj21}, SN 2020hvf \citep{jian21}, SN 2020esm \citep{dimi22}, SN 2021zny \citep{dimi23}, SN 2022ilv \citep{sriv23}.  The C\,{\sc ii} lines seen around $\sim -$10\,d in PSN J0910+5003 and ASASSN-16ex faded speedily, similar to SN 2006gz. The origin of carbon is not fully understood. The strong C\,{\sc ii} lines in the 09dc--like SNe suggest a large amount of unburned material in the SN ejecta. This supports the model with  an extended envelope.  SN 2020esm had a nearly pure C/O atmosphere during the first few days of the explosion. The large amount of carbon  in the envelope is  swept up by the ejecta giving rise to strong/persistent C\,{\sc ii} features \citep{taub13,noeb16,dimi22}. \citet{maed23} suggest that the sequence of SN 2020hvf,
2012dn and 2020esm/2009dc could be connected to an increase in the C/O-rich envelope, resulting in increasing strength of the C\,{\sc ii} and decreasing velocity of Si\,{\sc ii} as found by \citet{asha21} for a sample of 09dc--like SNe Ia.

The presence of carbon may also help in dust formation in these objects \citep{maed23}. 
Pre-existing dust or CO/dust formation may cause a fast decline in the light curves \citep{taub13,taub19,chak14,yama16,naga17,hsia20}. 
Early fast decline, starting within $\sim$ 1 -- 3 months from $B$ maximum  was seen in SN 2012dn \citep{chak14,yama16,taub19}, ASASSN-15pz \citep{chen19} and LSQ14fmg \citep{hsia20}. While a fast decline in the late phase was seen in SN 2006gz \citep{maed09}, SN 2009dc \citep{silv11,taub13} and SN 2020esm \citep{dimi22}. 
No signature of fast decline is seen in the light curves of PSN J0910+5003 and ASASSN-16ex till the last available data point ($\sim$ 4 months).  

Shell/envelope configuration helps to explain the low ejecta velocity and slow evolution \citep{noeb16}. 
Most of the 09dc--like SNe Ia, e.g., SN 2003fg, 07if, 09dc, 20esm, ASASSN-15pz, 15hy  were found to have very low ejecta velocities. Their velocities near the maximum phase are below 10\,000\,km\,s$^{-1}$. While SNe 2006gz, 12dn and LSQ14fmg were found to have higher velocities (refer  Fig.~\ref{fig_vel}). Both PSN J0910+5003 and ASASSN-16ex have velocity evolution similar to SN 2006gz. 
Along with high-velocity features, SN 2020hvf had a relatively high ejecta velocity. \citet{jian21} suggest that there could be different origins of 09dc--like SNe with different velocities.

All the 09dc--like SNe Ia can be placed in the normal velocity (NV) subgroup of \cite{wang09hv}. They have a lower gradient in their velocity evolution, falling in the LVG subgroup of \cite{bene05} classification scheme. PSN J0910+5003 and ASASSN-16ex both fall in NV and LVG subgroups. 
The  strength of Si\,{\sc ii}\,$\lambda$5972,  $\lambda$6355 lines in  09dc--like SNe Ia is found to be weak. The relative strength of these lines is temperature-sensitive \citep{nuge95}. Hotter events have a weaker Si\,{\sc ii}\,$\lambda$5972 line. At lower temperatures, Fe\,{\sc ii}/Co\,{\sc ii} line blanketing increases the strength of Si\,{\sc ii}\,$\lambda$5972, while at higher temperatures, Fe\,{\sc iii}/Co\,{\sc iii} line blanketing washes out this feature. 09dc--like SNe are UV blue and bright during the early/maximum phase, making their photosphere hotter, giving a weaker  Si\,{\sc ii}\,$\lambda$5972 line. The 09dc--like SNe mostly fall in the `shallow silicon' (SS) subgroup of the \cite{bran06} classification.
The Si\,{\sc ii}\,$\lambda$5972, $\lambda$6355 lines appear weak in the maximum phase spectrum of ASASSN-16ex; hence spectroscopically, it can be categorized in the SS subgroup. While PSN J0910+5003 falls near the boundary of the SS and CN subgroups. The velocity evolution of SN 2013bz is similar to SN 2003du. The Si\,{\sc ii}\,$\lambda$5972 line is shallower, falling in the `shallow silicon' (SS) subgroup. It has a normal velocity (NV) at maximum and a low-velocity gradient (LVG) in post-maximum phase. 

In this work, we have presented detailed photometric and spectroscopic analyses of the three supernovae: SN 2013bz, PSN J0910+5003 and ASASSN-16ex. SN 2013bz is a slow-declining, luminous event with photometric/spectral characteristics similar to normal SNe Ia. While most of the photometric/spectroscopic properties of PSNJ 0910+5003 and ASASSN-16ex are similar to 09dc--like SNe Ia. The combined data of these objects allow us to explore and compare various astrophysical parameters of normal and peculiar SNe Ia and  help understand differences in their characteristics. Adding well-studied new members to the family of peculiar SNe Ia will enhance our knowledge of these objects and SNe Ia diversity.

\section*{Acknowledgements}
We thank the anonymous referee for providing constructive comments, which improved the presentation of this paper.
ST and NKC are thankful to the Director and Dean of IIA, Bengaluru, for the local hospitality and facilities provided. We are thankful to the staff at CREST and IAO for their assistance during the observations and to all the observers of the 2-m HCT (IAO-IIA), who kindly provided part of their observing time for supernova observations. We have used public data in the {\it Swift} data archive. 
This work has made use of the NASA Astrophysics Data System (ADS), NASA/IPAC Infrared Science Archive (IRSA) and the NASA/IPAC Extragalactic Database (NED), which is operated by Jet Propulsion Laboratory, California Institute of Technology, under contract with the National Aeronautics and Space Administration. We acknowledge the use of Weizmann Interactive Supernova Data Repository (WISeREP) maintained by the Weizmann Institute of Science computing center.

%%%%%%%%%%%%%%%%%%%%%%%%%%%%%%%%%%%%%%%%%%%%%%%%%%
\section*{Data Availability}
The photometric and spectroscopic data  presented in this paper will be made available by the corresponding author upon request.

%%%%%%%%%%%%%%%%%%%% REFERENCES %%%%%%%%%%%%%%%%%%

% The best way to enter references is to use BibTeX:

\bibliographystyle{mnras}
\bibliography{Ref_list}

% Don't change these lines
\bsp	% typesetting comment
\label{lastpage}
\end{document}